\documentclass[aps,pre,twocolumn,superscriptaddress]{revtex4}
\usepackage[dvips]{graphics}
\usepackage{graphicx}
\usepackage{amsfonts}
\usepackage{amssymb}
\usepackage{amsmath}
\usepackage{color}

\begin{document}

\title{Existence, Stability, and Scattering of Bright Vortices
in the Cubic-Quintic Nonlinear Schr{\"o}dinger Equation}

\author{R.M.~Caplan}

\author{R.\ Carretero-Gonz\'alez}
\affiliation{Nonlinear Dynamical Systems Group,%
\footnote{\texttt{URL}: http://nlds.sdsu.edu}%
Department of Mathematics and Statistics, and Computational Science
Research Center, San Diego State University, San Diego CA, 92182-7720, USA}

\author{P.G.\ Kevrekidis}
\affiliation{Department of Mathematics and Statistics, University of Massachusetts,
Amherst MA 01003-4515, USA}

\author{B.A.~Malomed}
\affiliation{Department of Physical Electronics, Faculty of Engineering,
Tel Aviv University, Tel Aviv 69978, Israel}

\date{\today}

\begin{abstract}
We revisit the topic of the existence and azimuthal
modulational stability of solitary vortices (alias
vortex solitons) in the two-dimensional (2D) cubic-quintic
nonlinear Schr{\"o}dinger equation. We develop a
semi-analytical approach, assuming that the vortex
soliton is relatively narrow, and thus splitting
the full 2D equation into radial and azimuthal 1D
equations. A variational approach is used to
predict the radial shape of the vortex soliton, using
the radial equation, yielding results very close to those obtained
from numerical solutions. Previously known existence bounds
for the solitary vortices are recovered by means of this
approach. The 1D azimuthal equation of motion is used to analyze the
modulational instability of the vortex solitons. The semi-analytical predictions -- in particular,
that for the critical intrinsic frequency of the vortex
soliton at the instability border -- are compared to
systematic 2D simulations. We also compare our
findings to those reported in earlier works, which
featured some discrepancies.
We then perform a detailed computational study of collisions 
between stable vortices with different topological charges. Borders
between elastic and destructive collisions are identified.
\end{abstract}


\maketitle

\section{Introduction}

\label{sec:intro}

The cubic-quintic nonlinear Schr{\"{o}}dinger (CQNLS) equation is used to
model a variety of physical settings. In scaled units, the CQNLS equation
takes the following form:
\begin{equation}
i\frac{\partial \Psi }{\partial t}+\nabla ^{2}\Psi +\left\vert \Psi
\right\vert ^{2}\!\Psi -\left\vert \Psi \right\vert ^{4}\!\Psi =0,
\label{CQNLS}
\end{equation}%
where $\Psi (x,y,t)$ is the complex wave function, $\nabla ^{2}$ is the
two-dimensional (2D) Laplacian, and the last two terms represent,
respectively, the focusing cubic and defocusing quintic nonlinearities. The
CQNLS equation emerges in models of light propagation in diverse optical
media, such as non-Kerr crystals \cite{PTS}, chalcogenide glasses \cite%
{CQglass1,CQglass2}, organic materials \cite{CQorganic}, colloids \cite%
{colloid1,colloid2}, dye solutions \cite{dye}, and ferroelectrics \cite%
{ferroelectric}. It has also been predicted that this complex nonlinearity
can be synthesized by means of a cascading mechanism \cite{cascading}. It
should be noticed that, in the optics models, evolution variable $t$ is not
time, but rather the propagation distance.

\begin{figure}[tbh]
\begin{center}
\includegraphics[width=3.0in]{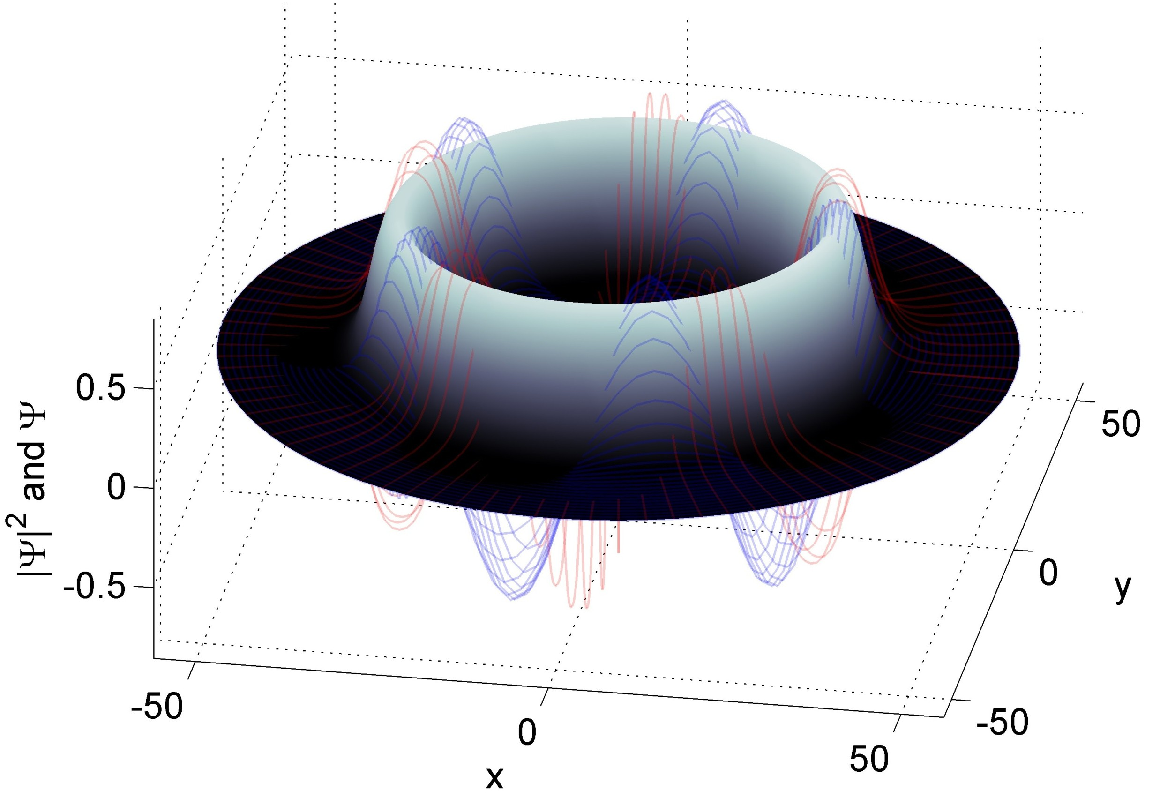}
\end{center}
\vspace{-0.3cm}
\caption{(Color online) An example of a vortex soliton with charge $m=5$ and
frequency $\Omega =0.16$, found as a numerical solution to the CQNLS
equation. The height of the profiles represents the value of the wave
function. The gray profile displays the squared absolute value, while the
dark (blue) and light (red) meshes represent the real and imaginary parts,
respectively. }
\label{v}
\end{figure}

The competition of the focusing (cubic) and defocusing (quintic) nonlinear
terms is the key feature of the CQNLS model, which allows for the existence
of stable multidimensional structures which would be unstable in the
focusing cubic nonlinear Schr{\"{o}}dinger (NLS) equation. In particular,
the CQNLS equation supports solitary vortices (alias \textit{vortex solitons}%
) in 2D and 3D geometries. These are ring-shaped structures which carry a
rotational angular phase. The respective integer winding number in the
vortex is referred to as its topological charge (alias \textit{vorticity}), $m$. An example of a stable solitary vortex is depicted in Fig.~\ref{v}.

\begin{figure}[tbh]
\begin{center}
\includegraphics[width=3.5in]{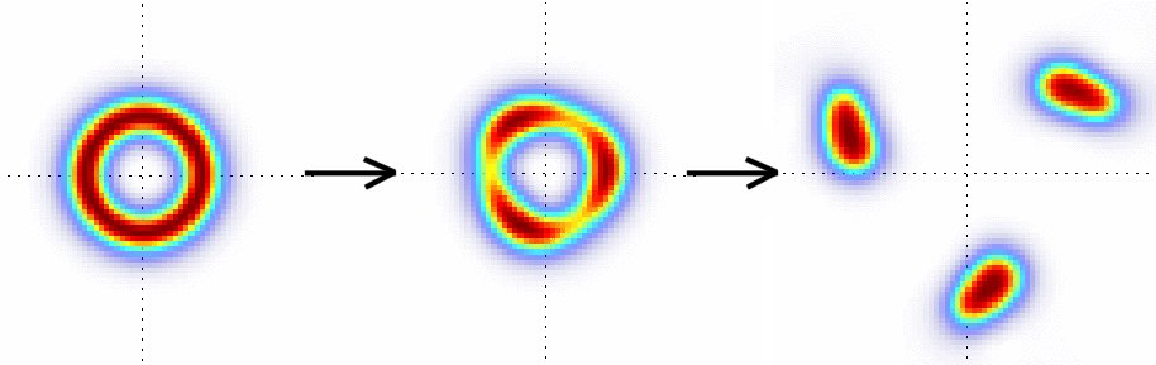}
\end{center}
\vspace{-0.3cm}
\caption{(Color online) An example of the evolution of the vortex soliton
with charge $m=4$ in the CQNLS equation, exhibiting breakup (left to
right) due to the azimuthal modulational instability. Shown is the intensity
(squared absolute value of the wave function) at consecutive moments of
time, from left to right. The numerical method and parameters are presented
in Sec.~\protect\ref{sec:AMSresults}.}
\label{MI}
\end{figure}

It is well known that the cubic NLS equation supports ``dark" (delocalized) and ``bright" (localized) vortices,
with the self-defocusing and focusing signs of the nonlinearity,
respectively. Dark vortices of unitary charge are stable,
while higher-order ones are unstable, splitting into unitary eddies \cite%
{Fetter}. On the other hand, the ``bright" modes (vortex solitons)
are always azimuthally unstable in the framework of the cubic
equation or its counterpart with the saturable nonlinearity
\cite{Skryabin},
which leads to the breaking of the vortex soliton into a number of fragments depending on the charge \cite%
{NLSxMI,Torner}. The addition of a periodic potential to the cubic NLS
equation may stabilize vortex solitons of a different type, which are built
(in the simplest case) as a chain of four peaks with the phase circulation
corresponding to integer vorticity $m$
\cite{pu99,Baizakov,Yang,herring08,Thawatchai}.
An example of the azimuthal breakup of an unstable vortex is displayed in Fig.~\ref{MI}.

In contrast to the cubic NLS, the CQNLS can support \emph{stable} solitary vortices. For the first time, this
remarkable fact was discovered ``empirically" in simulations of the
2D CQNLS equation \cite{AMIxCQNLSxOLD}, and later was investigated
in detail in a more rigorous form in
Refs.~\cite{AMIxCQNLSxSHORT,AMIxCQNLSxLONG}. Moreover,
\emph{three-dimensional} solitons with embedded vorticity $m=1$ also
have their stability region in the 3D CQNLS equation\cite{PRL}
(see also reviews \cite{AMIxCQNLSxMID,Torner}). Stable
vortex solitons may find their potential applications in the design
of all-optical data-processing schemes. In that respect, knowing the
growth rates of unstable modes is also important, because, if the
rates are small enough, the vortices may be considered as
practically stable ones, as they will not exhibit an observable
instability over relevant propagation distances.

One of objectives of this paper is to revisit the topic of the azimuthal
modulational stability of solitary-vortex solutions in the 2D CQNLS
equation. As mentioned above, numerous studies of this problem have already
been published \cite%
{AMIxCQNLSxOLD,AMIxCQNLSxSHORT,AMIxCQNLSxMID,AMIxCQNLSxLONG,AMIxCQNLSxNEW}.
The aim of the analysis was to find a critical value ($\Omega _{%
\mbox{\scriptsize
st}}$) of the intrinsic frequency of the vortex [for its exact definition
see Eqs.~(\ref{var_sep}) and (\ref{A}) below], above which the vortices with
a given value of the topological charge are \emph{stable}. In Ref.~\cite%
{AMIxCQNLSxOLD}, 2D azimuthally stable vortices with $m=1$ were shown to
exist. It was found that the slope of the vortex' profile at the pivotal
point peaked at a specific value of $\Omega $, which was considered as the
critical frequency, $\Omega _{\mbox{\scriptsize st}}(m=1)\approx 0.145$,
while the vortices, with all values of $m$, exist at $\Omega <\Omega _{%
\mbox{\scriptsize max}}^{\mbox{\scriptsize 1D}}=3/16\equiv 0.1875$ (at $%
\Omega =\Omega _{\mbox{\scriptsize max}}^{\mbox{\scriptsize 1D}}$,
the radius of the vortex diverges, i.e., the ``bright" vortex
goes over into a ``dark" one). The same value, $\Omega _{%
\mbox{\scriptsize max}}^{\mbox{\scriptsize 1D}}$ is simultaneously the
largest one up to which exact soliton solutions exist in the 1D version of
the CQNLS equation \cite{Bulgaria}. In work \cite{AMIxCQNLSxOLD}, a
variational approach (VA)\ was developed for 2D vortex solitons, yielding
results similar to those obtained in the numerical form. Full 2D simulations
of the CQNLS equation reported in Ref. \cite{AMIxCQNLSxOLD} had confirmed
that vortices with $\Omega _{\mbox{\scriptsize st}}<\Omega <\Omega _{%
\mbox{\scriptsize max}}^{\mbox{\scriptsize 1D}}$ were indeed stable, while
those with $0<\Omega <\Omega _{\mbox{\scriptsize st}}$ were not.

Works~\cite{AMIxCQNLSxSHORT,AMIxCQNLSxMID,AMIxCQNLSxLONG} employed a more
rigorous approach. They introduced small perturbations around the 2D vortex
soliton and solved the resulting eigenvalue problem numerically. In this
way, the critical frequencies were found as $\Omega _{\mbox{\scriptsize st}%
}(m=1)\approx 0.16$ and $\Omega _{\mbox{\scriptsize st}}(m=2)\approx 0.17$.
Also, in Ref.~\cite{Victor08}, using the Gagliardo-Nirenberg and H\"{o}lder
inequalities together with Pohozaev identities, is was shown that the
eigenvalues generated by the CQNLS equation possess an upper cutoff value.

In Ref.~\cite{AMIxCQNLSxNEW}, 2D perturbations were considered too. Through
an extensive analysis, the problem was transformed into finding, by means of
numerical methods, zeros of respective Evans functions for a set of ordinary
differential equations (ODEs). In this way, the existence of azimuthally
stable vortices of all integer values of $m$ was predicted (in Refs.~\cite%
{AMIxCQNLSxSHORT} and \cite{AMIxCQNLSxOLD}, the stability regions for $m\geq
3$ were not identified, as they are extremely small). The predictions for
the critical frequencies reported in Ref. \cite{AMIxCQNLSxNEW}, which are
somewhat different from those in Refs.~\cite{AMIxCQNLSxSHORT} and \cite%
{AMIxCQNLSxOLD}, are shown in Table~\ref{t:results}.

The present work aims to undertake an additional study of the stability of
vortices in the CQNLS equation for two reasons. Firstly, the previous
studies demonstrated some (relatively small, but not negligible)
disagreement in the predictions of the critical frequency. Therefore, since
we will use a different approach, our results may help in comparing and
verifying the previous findings. Secondly, although in Ref.~\cite%
{AMIxCQNLSxOLD} 2D simulations were performed for the vortices, this was
only done for $m=1$, and longer simulations later showed that some vortices
which were originally concluded to be stable turned out to be eventually
unstable \cite{AMIxCQNLSxSHORT}. Therefore, stability results from additional 2D numerical simulations are needed for comparison with various
predictions of the stability. In fact, our numerical results will illustrate
that the predictions of Ref.~\cite{AMIxCQNLSxNEW} are the most accurate
ones. In parallel to extensive direct simulations, we elaborate a
semi-analytical methodology, following the lines of previous studies of the
azimuthal modulational instability of vortices in the cubic NLS
equation \cite{NLSxMI}. The approximation amounts to splitting the full 2D
equation into effectively one-dimensional radial and azimuthal equations.
The former equation is used to compute the radial shape of the soliton by means of a
VA to find an analytic approximation to the profile, which is then used as an initial condition to a numerical optimization routine to find the numerically `exact' profile.  The closeness between the numerically computed profiles and the VA approximation testify to the good accuracy of the VA.  The analysis of the azimuthal equation makes
it possible to predict the threshold of the onset of the splitting
instability of the vortex soliton. Our results predicting the growth
rates of individual azimuthal modes of unstable vortices match full
simulations very well, but the predictions for the critical frequency are
less accurate, when compared to the simulations. This semi-analytical
technique may be quite relevant for applications in other 2D models.

Another objective of the work is to study, by means of systematic
simulations, collisions between stable vortices. Except for few examples
reported in Ref. \cite{AMIxCQNLSxOLD}, this problem was not studied before.

The paper is organized as follows. In Sec.~2 we derive an approximate
analytical description of the vortex profile by first finding an analytical
asymptotic approximation to it, which is then employed as the \textit{ansatz}
on which the VA is based. In Sec.~3 we use the variational ansatz as the
initial guess, to generate numerically exact vortex profiles by dint of a
nonlinear optimization routine. We then compare the numerical profiles with
the variational ansatz to show its accuracy. In Sec.~4 we derive an
approximate 1D equation for the dynamics along the azimuthal direction. A
linear stability analysis is subsequently performed to find stability
criteria and growth rates of unstable modes within the framework of the
azimuthal equation. In Sec.~5 we present full 2D simulations of the
vortices, and the respective results for their stability. These results are
compared to our predictions, as well as to the predictions produced by
previous works. In Sec.~6, we use direct simulations to explore collisions
between stable vortices in detail. In particular, these studies allow us to
identify a border between quasi-elastic and destructive collisions. Finally,
in Sec.~7, we summarize our findings and formulate concluding remarks.

\section{Approximate Analytical Profiles of Steady-State Vortices}

\label{sec:VA}

A steady-state vortex solution to Eq.~(\ref{CQNLS}) is looked for as
\begin{equation}
\Psi (r,\theta ,t)=f(r)\,A(\theta ,t),  \label{var_sep}
\end{equation}%
where real function $f(r)$ is a stationary radial profile with azimuthal
dependence given by
\begin{equation}
A(\theta ,t)=e^{i(m\theta +\Omega t)},  \label{A}
\end{equation}%
with topological charge $m$ and frequency $\Omega $. Analytical solutions
being not available for the profile $f(r)$, we begin the study by finding an
approximate analytic expression for $f(r)$ and identifying its existence
bounds. As shown below in Sec.~\ref{sec:NOresults}, the predicted profile is
very close to the true solution, therefore it can be used to predict the
existence and stability regions of the vortex solitons. The developed
analytical method is quite general and may be applied to other physically
relevant equations.

\subsection{The Asymptotic Vortex Profile}

\label{2DCQNLSprofiles}

Inserting expression~(\ref{var_sep}) into the underlying equation (\ref%
{CQNLS}) yields the following ODE for the radial profile of the vortex with
charge $m$:
\begin{equation}
\frac{1}{r}\frac{d}{dr}\left( r\frac{df}{dr}\right) -\left( \Omega +\frac{%
m^{2}}{r^{2}}\right) f(r)+f^{3}(r)-f^{5}(r)=0.  \label{SepCQNLS}
\end{equation}
If we assume that the vortex takes the form of a relatively narrow ring with
a large radius, then, in the region of interest, variable $r$ may be
approximately replaced by a constant, $r_{c}\gg 1$, which we take to be the
value of $r$ at a point where the radial profile attains its maximum. Since
we assume $r_{c}$ to be large, in the lowest approximation we neglect the $1/r_{c}$
term in the Laplacian. An obvious consequence of dropping this term is that
the approximate profiles which we are going to derive will be less accurate
for smaller values of $r_{c}$, see Sec.~\ref{sec:NOresults}.

Under the large-radius assumption, Eq.~(\ref{SepCQNLS}) becomes
\begin{equation}
\frac{d^{2}\!f}{dr^{2}}-\Omega ^{\ast }\,f(r)+f^{3}(r)-f^{5}(r)=0,
\label{SepCQNLSOMstar}
\end{equation}
\begin{equation}
\Omega ^{\ast }\equiv \Omega +m^{2}/r_{c}^{2}.  \label{omstar}
\end{equation}
For a relatively narrow ring, one can treat Eq.~(\ref{SepCQNLSOMstar}) as if
$r\in (-\infty ,+\infty )$, in which case the equation has a well-known 1D
analytical solution for the soliton \cite{Bulgaria,CQNLS1DSOL}:
\begin{equation}
f_{\mbox{\scriptsize asy}}^{2}(r)=\frac{4\Omega ^{\ast }}{1+\sqrt{%
1-(16/3)\Omega ^{\ast }}\,\cosh \left( 2\sqrt{\Omega ^{\ast }}%
\,(r-r_{c})\right) }.  \label{2dcqnls_sol}
\end{equation}%
As mentioned above, this solution exists for $0<\Omega ^{\ast }<\Omega _{%
\mbox{\scriptsize max}}^{\mbox{\scriptsize
1D}}\equiv 3/16$. At point $\Omega ^{\ast }=3/16$, the solution degenerates
into a constant, $f(r)=f_{0}\equiv \sqrt{3/4}$. The family of solutions (\ref%
{2dcqnls_sol}) parameterized by $\Omega ^{\ast }$ is depicted in Fig.~\ref%
{1DCQNLS_sols}.

\begin{figure}[tbh]
\begin{center}
\includegraphics[width=3.5in]{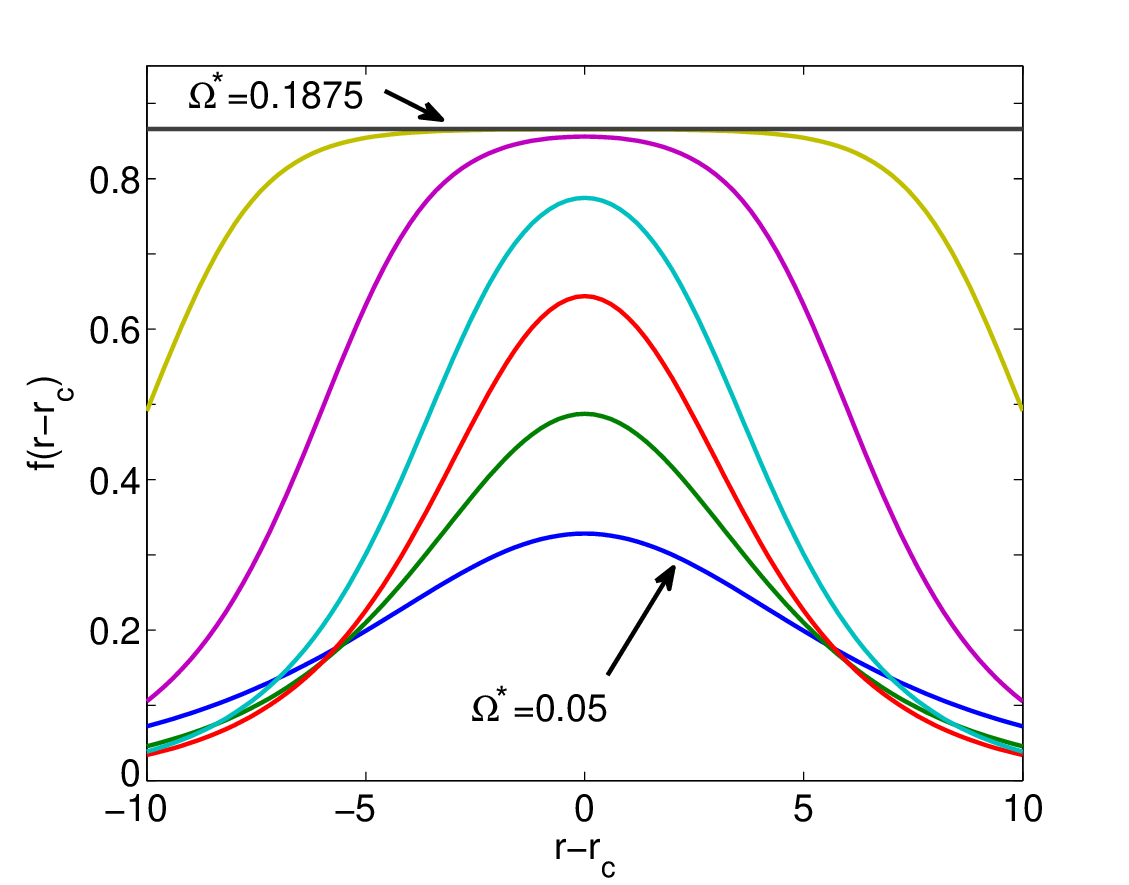}
\end{center}
\vspace{-0.3cm}
\caption{(Color online) A set of solutions (\protect\ref{2dcqnls_sol}) for $%
\Omega ^{\ast }=0.05,0.1,0.15,0.18,0.1874,0.1874999,0.1875$ are shown from
bottom to top. As $\Omega ^{\ast }$ increases, the profile flattens out, and
it degenerates into a constant at $\Omega =\Omega ^{\ast }$.}
\label{1DCQNLS_sols}
\end{figure}
To obtain a closed-form analytical approximation, it is now necessary to
find an expression for the radius of the profile, $r_{c}$, for given
frequency $\Omega $. To this end, we will use the VA, and will then compare
the results with numerically found profiles.

\subsection{The Variational Approach}

\label{sec:va}

To apply the VA, we use the Lagrangian density corresponding to the CQNLS
equation:
\begin{equation}
\mathcal{L}=\frac{i}{2}\left( \Psi \Psi _{t}^{\ast }-\Psi ^{\ast }\Psi
_{t}\right) +|\Psi _{r}|^{2}+r^{-2}|\Psi _{\theta }|^{2}-\frac{1}{2}|\Psi
|^{4}+\frac{1}{3}|\Psi |^{6},  \label{Ldence}
\end{equation}%
where the subscripts denote partial derivatives. Inserting here expression (%
\ref{var_sep}), to be used as a factorized \textit{ansatz}, yields
\begin{equation}
\mathcal{L}=\left( \Omega +\frac{m^{2}}{r^{2}}\right) f^{2}(r)+\left( \frac{%
df}{dr}\right) ^{2}-\frac{1}{2}f^{4}(r)+\frac{1}{3}f^{6}(r).  \label{var_lag}
\end{equation}%
The integration of density (\ref{var_lag}) gives rise to the full Lagrangian:%
\begin{equation}
L\equiv 2\pi \int_{0}^{\infty }\mathcal{L}(r)dr=2\pi \left( \Omega
\,C_{1}+C_{2}+m^{2}\,C_{3}-\frac{1}{2}\,C_{4}+\frac{1}{3}\,C_{5}\right) ,
\label{var_langC}
\end{equation}
where
\begin{alignat}{3}
C_{1}& \equiv \int_{0}^{\infty }f^{2}(r)r\,dr, 
& \qquad  \notag
C_{2}& \equiv \int_{0}^{\infty }\left( \frac{df}{dr}\right) ^{2}r\,dr,
\\[1.0ex]
C_{3}& \equiv \int_{0}^{\infty }\!f^{2}(r)\,\frac{dr}{r},  
& \qquad 
C_{4}& \equiv \int_{0}^{\infty }\!f^{4}(r)r\,dr, 
\label{constants} \\[1ex]
C_{5}& \equiv \int_{0}^{\infty }\!f^{6}(r)r\,dr.
& \qquad   \notag
\end{alignat}

Next, we use the asymptotic approximate profile (\ref{2dcqnls_sol}) as an
\textit{ansatz} for radial profile, treating $\Omega ^{\ast }$ and $r_{c}$
in Eq.~(\ref{2dcqnls_sol}) as variational parameters. Inserting expression~(%
\ref{2dcqnls_sol}) into the integral terms in the Lagrangian, and again
making use of the large-radius approximation yields%
\begin{align}
C_{1}& \approx 2\sqrt{3}\,T\,r_{c},\quad C_{2}\approx r_{c}\left[ \frac{3}{8%
}\sqrt{\Omega ^{\ast }}-\sqrt{3}\,T\,\left( \frac{3}{16}-\Omega ^{\ast
}\right) \right],
\notag
\\[1ex]
 \label{cconstsva}
C_{3}& \approx \frac{2\sqrt{3}\,T}{r_{c}},\quad C_{4}\approx -3\,r_{c}\left[
\sqrt{\Omega ^{\ast }}-\frac{\sqrt{3}}{2}\,T\right] , \\[1ex]
C_{5}& \approx -r_{c}\left[ \frac{27}{8}\sqrt{\Omega ^{\ast }}-3\sqrt{3}%
\,T\left( \frac{9}{16}-\Omega ^{\ast }\right) \right] ,  \notag
\end{align}
where
\begin{equation}
T\equiv \mbox{arctanh}\left[ \sqrt{\frac{3}{16\Omega ^{\ast }}}-\sqrt{\frac{3%
}{16\Omega ^{\ast }}-1}\right] .  \label{T}
\end{equation}%
The respective static Euler-Lagrange equations, ${\partial L}/{\partial r_{c}%
}=0$ and ${\partial L}/{\partial \Omega ^{\ast }}=0$, yield, respectively,
equation%
%
\begin{equation}
r_{c}=m\left( \Omega -\frac{3}{16}+\frac{\sqrt{3\Omega ^{\ast }}}{8T}\right)
^{-1/2},  \label{var_eullan2}
\end{equation}
and the other one which is equivalent to Eq.~(\ref{omstar}). Eliminating $%
r_{c}$ from these equations, one concludes that $\Omega ^{\ast }$ is, for a
chosen value of $\Omega $, a solution to the following equation,
\begin{equation}
\Omega =\frac{\Omega ^{\ast }}{2}+\frac{3}{32}-\frac{\sqrt{3\Omega ^{\ast }}%
}{16T}\equiv G(\Omega ^{\ast }).  \label{omfunction}
\end{equation}%
This yields the final form of the vortex radial profile, as predicted by the
VA:
\begin{equation}
f_{\mbox{\scriptsize va}}^{2}(r)=\frac{4\Omega ^{\ast }}{1+\sqrt{%
1-(16/3)\Omega ^{\ast }}\,\cosh \left( 2\sqrt{\Omega ^{\ast }}\,(r-r_{c}^{%
\mbox{\scriptsize va}})\right) },  \label{VAeq}
\end{equation}%
\begin{equation}
r_{c}^{\mbox{\scriptsize va}}\equiv \frac{m}{\sqrt{\Omega ^{\ast }-G(\Omega
^{\ast })}}.  \label{rcoms}
\end{equation}
Finally, the transcendental equation~(\ref{omfunction}), together with
Eq.~(\ref{T}), was solved numerically by means of a simple bisection method with a tolerance of $10^{-15}$.

\begin{figure}[tbh]
\begin{center}
~~~~~~\includegraphics[width=2.75in]{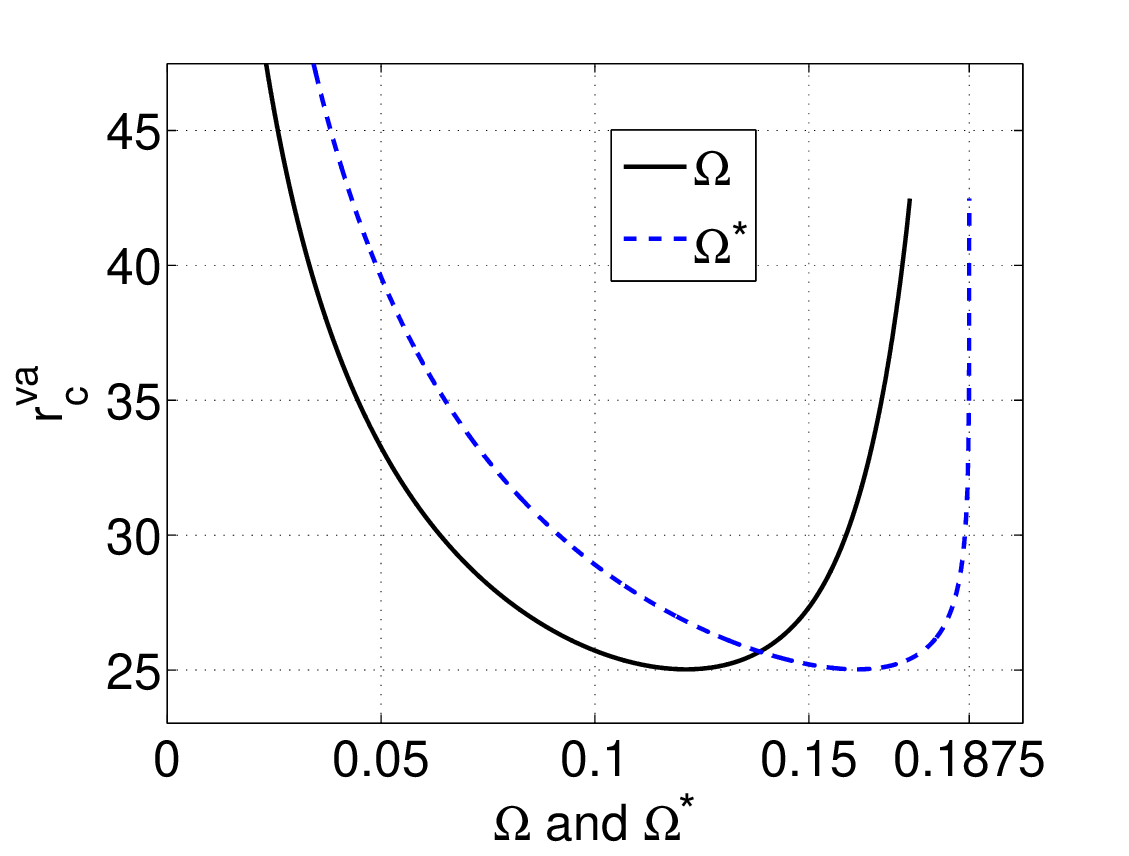} 
\end{center}
\begin{center}
\includegraphics[width=2.75in]{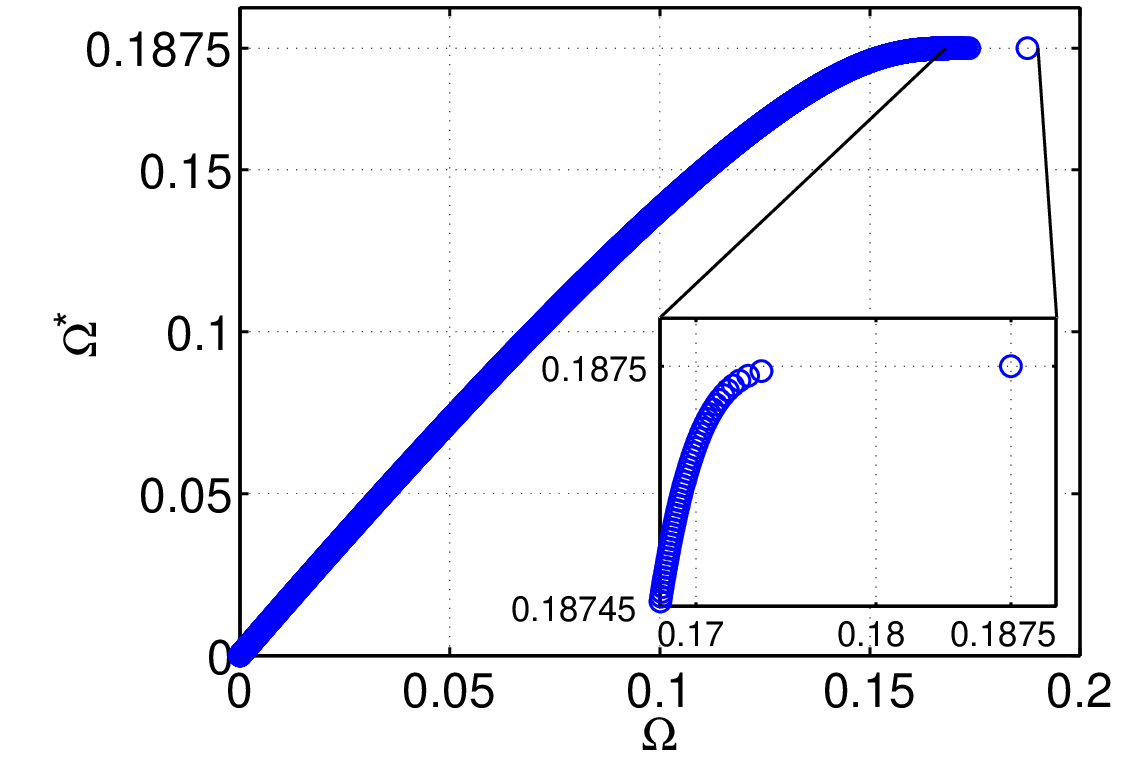}
\end{center}
\vspace{-0.3cm}
\caption{(Color online) Top: The vortex' radius, as predicted by the
variational approximation, $r_{c}^{\mbox{\scriptsize va}},$ versus $\Omega
=G(\Omega ^{\ast })$, and $\Omega ^{\ast }$, for $m=5$. 
Bottom: The values of
$\Omega =G(\Omega ^{\ast })$ vs. $\Omega ^{\ast }$, see Eq.~(\protect\ref%
{omstar}). $G(\Omega ^{\ast })$ was evaluated over interval $\Omega ^{\ast
}\in \left[ 0.01,0.1875\right] $ with step $\Delta \Omega ^{\ast }=0.00001$.
}
\label{rovaomom}
\end{figure}

In Fig.~\ref{rovaomom} we show $r_{c}^{\mbox{\scriptsize va}}$ versus $%
\Omega $ and $\Omega ^{\ast }$ for $m=5$, along with relationship (\ref%
{omfunction}). We see that, with the increase of $\Omega $, the radius of
the vortex starts from very large values (infinite at $\Omega =0$),
decreases until it reaches a minimum value, and then increases rapidly (it
becomes infinite once again at $\Omega =\Omega _{\mbox{\scriptsize max}}^{%
\mbox{\scriptsize 2D}}$). We also see that the relationship between $\Omega $
and $\Omega ^{\ast }$ does not depend on charge $m$, and that $\Omega
\rightarrow 3/16\ $as $\Omega ^{\ast }\rightarrow 0.1875$ (the apparent gap
near the right edge of the plot is due to the sensitivity of the
relationship, which we discuss below).

\subsection{Existence Bounds for the Two-Dimensional Vortex Profiles}

\label{sec:ebounds}

From numerical investigations performed in Refs.~\cite{AMIxCQNLSxOLD} and
\cite{AMIxCQNLSx2D3D}, the existence border for the two-dimensional CQNLS
equation vortex solution was found to be $\left( \Omega _{%
\mbox{\scriptsize
max}}^{\mbox{\scriptsize 2D}}\right) _{\mathrm{num}}\approx 0.180$, while,
as said above, the analytical limit, which is identical for 1D and 2D
equations, is $\Omega _{\mbox{\scriptsize max}}^{\mbox{\scriptsize 2D}%
}=\Omega _{\mbox{\scriptsize max}}^{\mbox{\scriptsize 1D}}=3/16=0.1875.$ The
discrepancy may be explained by the sensitivity of relationship (\ref%
{omfunction}) between $\Omega $ and $\Omega ^{\ast }$, as produced by the
variational approximation. From Eqs.~(\ref{omfunction}) and (\ref{T}) we see
that, as $\Omega ^{\ast }\rightarrow 3/16$, $T\rightarrow \infty $, and so $%
\Omega =G(\Omega ^{\ast })\rightarrow \Omega ^{\ast }/2+3/32=3/16$, in
accordance with the exact result. However, as seen in the right panel of
Fig.~\ref{rovaomom}, the relationship between $\Omega $ and $\Omega ^{\ast }$
seems to be extremely sensitive near this limit (see the inset in the right
panel of Fig.~\ref{rovaomom}). This sensitivity can also be observed in
Table~\ref{t:omsen}, where some values of $\Omega ^{\ast }$ and the
corresponding $G(\Omega ^{\ast })=\Omega $ are given. It is clearly seen
that one quickly approaches the limit of machine double precision for $\Omega
^{\ast }$ as $\Omega \rightarrow 0.1875$.

\begin{table}[tbh]
\begin{center}
\begin{tabular}{|l|c|}
\hline
$\Omega^*$ & $\Omega = G(\Omega^*)$ \\ \hline\hline
0.1874 & 0.1664\dots \\ \hline
0.187499 & 0.1736\dots \\ \hline
0.187499999 & 0.1783\dots \\ \hline
0.187499999999 & 0.1806\dots \\ \hline
0.1874999999999999 & 0.1823\dots \\ \hline
\end{tabular}%
\end{center}
\vspace{-0.3cm}
\caption{The evaluation of $\Omega =G(\Omega ^{\ast })$ near $\Omega _{
\mbox{\scriptsize max}}^{\mbox{\scriptsize 2D}}=0.1875$ for double precision
arithmetic. Four significant digits in $G(\Omega ^{\ast })$ are given. The
extreme sensitivity of relationship $\Omega =G(\Omega ^{\ast })$ is
observed. To obtain more precise $\Omega ^{\ast }$ values for $\Omega $
closer to $\Omega _{\mbox{\scriptsize max}}^{\mbox{\scriptsize 2D}}=0.1875$,
the use of higher-precision arithmetic is required (results not shown here).
}
\label{t:omsen}
\end{table}

This effect can also be understood from the logarithmic divergence of $%
T(\Omega ^{\ast })$ close to the limit point, see Eq.~(\ref{T}). Therefore,
it is not surprising then that numerical estimates of $\Omega _{%
\mbox{\scriptsize
max}}^{\mbox{\scriptsize 2D}}$, obtained by means of a shooting method to
look for profiles at different vales of $\Omega $, gave lower estimates of
the existence bound. In Sec.~\ref{sec:NOresults}, we confirm, using accurate
numerical methods to find vortex profiles for $\Omega >0.180$, that the
actual existence bound $\Omega _{\mbox{\scriptsize max}}^{%
\mbox{\scriptsize
2D}}$ is greater than the numerical estimates of Refs.~\cite{AMIxCQNLSxOLD}
and \cite{AMIxCQNLSx2D3D}.

The analytical approximation (\ref{VAeq}) of the vortex 
profile is used in the following
sections as an input to a numerical optimization routine which solves the ODE (\ref{SepCQNLS}) to find numerically exact radial profiles.

\begin{figure*}[tbh]
\begin{center}
\includegraphics[height=5.5cm]{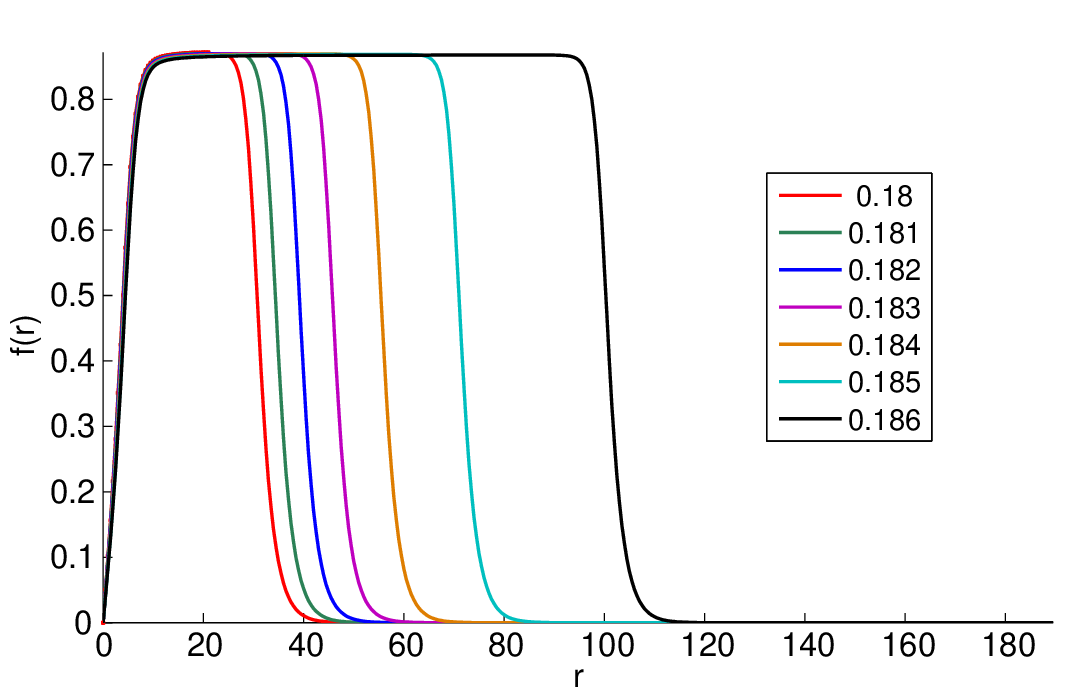} %
\includegraphics[height=5.5cm]{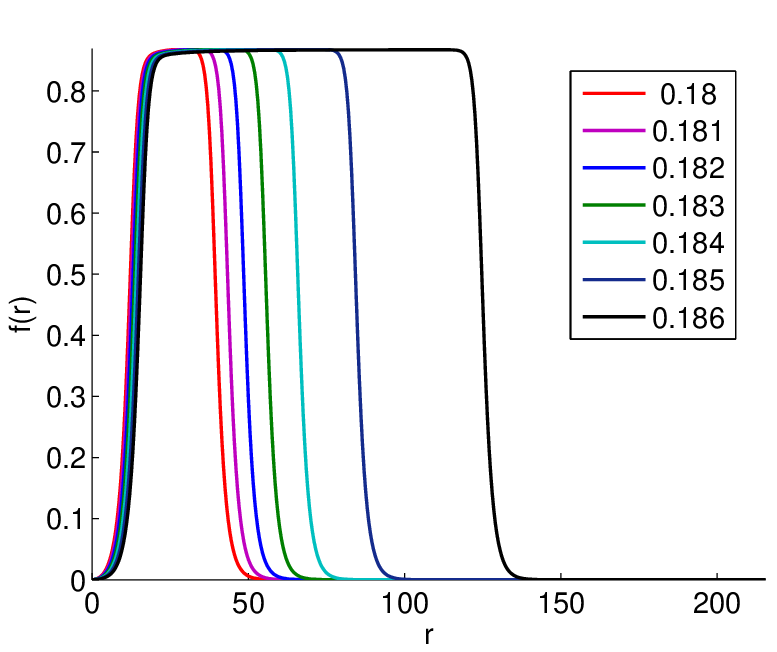}
\par
\includegraphics[height=6.25cm]{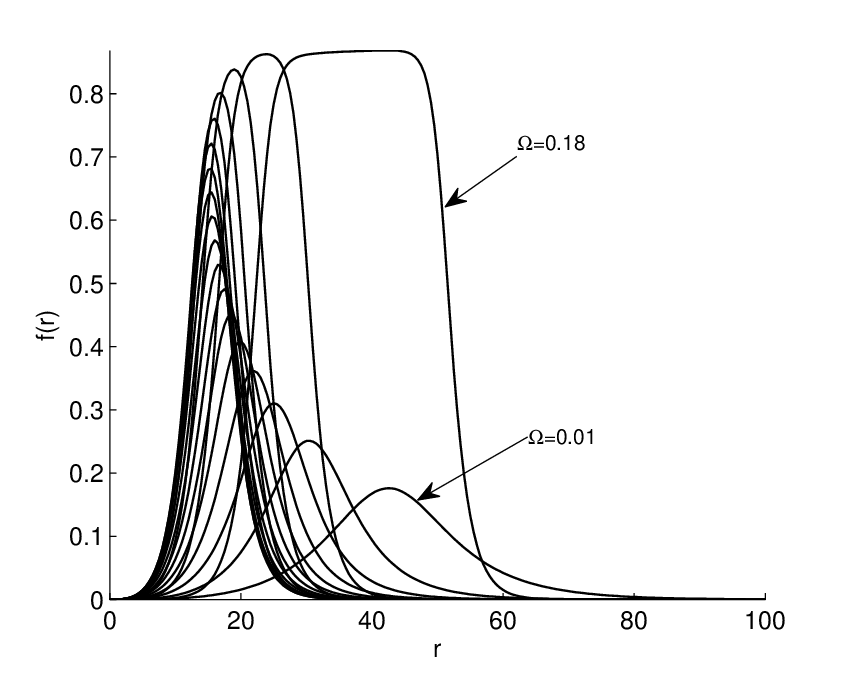} %
\includegraphics[height=6.25cm]{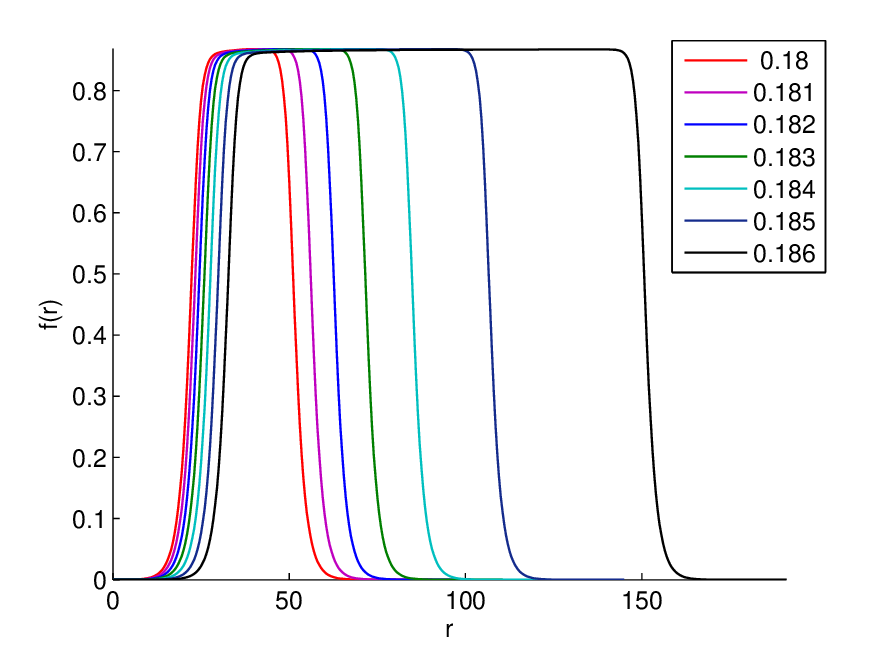}
\end{center}
\vspace{-0.3cm}
\caption{Steady-state vortex radial profiles computed by means of the
Gauss-Newton routine for $m=1$ (top left), $m=2$ (top right), and $m=3$ (the
bottom row of panels) for the indicated values of $\Omega $. In all cases,
the variational ansatz was used as the input (not shown here) with the value
of $\Omega ^{\ast }$ computed using high-precision arithmetic (see text for
details). The radial direction was discretized with a grid spacing of $%
\Delta r=0.5$. The stopping tolerance of $10^{-7}$ was used in the
Gauss-Newton routine. As $\Omega $ approaches the limit value, $3/16$, the
profiles flatten out due to the increase of the vortex' radius.}
\label{fig:profiles}
\end{figure*}

\section{Numerically Exact Steady-State Vortex Profiles and Comparison to
the Variational approximation}

\label{sec:NOresults}

To find numerical solutions to Eq.~(\ref{SepCQNLS}), we used a modified
Gauss-Newton optimization routine, with a tolerance $10^{-7}$ \cite{NLSxMI}.
In Fig.~\ref{fig:profiles}, some numerically found radial profiles are
displayed for charges $m=1$, $m=2$, and $m=3$ and $\Omega >0.18$, including $%
\Omega $ as close to $\Omega _{\mbox{\scriptsize max}}^{\mbox{\scriptsize 2D}%
}=\allowbreak 0.187\,5$ as $\Omega =0.186$. To our knowledge, the profiles above $\Omega = 0.181$ have not been shown in any previous study.  It is seen that the profiles
flatten out as $\Omega $ grows, pushing the profile farther from $r=0$. At
the same time, the ring of the vortex becomes wider without a change in its
height. It is worth mentioning that, due to the high sensitivity of the
numerics mentioned in the previous sections, computing profiles past $\Omega
=0.18$ is a daunting task requiring high-precision arithmetic. Implementing
high-precision arithmetics in a Gauss-Newton optimization routine (or any
fixed-point iteration method) would be quite involved and very time
consuming. Nonetheless, we have found that, using high-precision arithmetics
(up to $300$ decimal places for frequencies as close to $\allowbreak
0.187\,5 $ as $0.187$) in the calculation of $\Omega ^{\ast }$ in the
variational equation~(\ref{omfunction}) yields approximate analytical
solutions that the Gauss-Newton subroutine is able to process into
numerically accurate solutions for values of $\Omega $ very close to $\Omega
_{\mbox{\scriptsize
max}}^{\mbox{\scriptsize 2D}}=0.1875$, as shown in Fig.~\ref{fig:profiles}.
The results clearly demonstrate the \emph{usefulness of the VA}: without
using this approximation for generating the initial profiles to be fed into
the numerical solver, a prohibitively complex high-precision fixed-point
algorithm would be needed to obtain meaningful profiles past $\Omega =0.18$.

\begin{figure}[tbh]
\begin{center}
\includegraphics[width=3.5in]{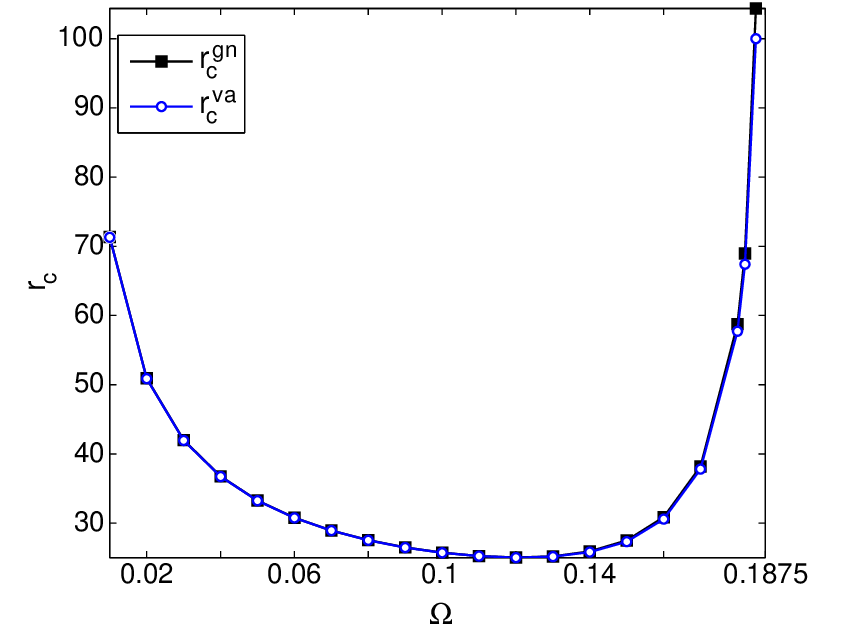}
\end{center}
\vspace{-0.3cm}
\caption{ The vortex-ring radius $r_{c}$ versus $\Omega $ as predicted by
the variational approximation ($r_{c}^{\mbox{\scriptsize va}}$, white
circles), and produced by the Gauss-Newton profile ($r_{c}^{
\mbox{\scriptsize gn}}$, black squares) for $m=5$. The numerical radius was
computed as the radial center of mass of the vortex ring.}
\label{rcresults}
\end{figure}

The accuracy of the variational ansatz per se can be tested by comparing it
to the numerical solutions computed by means of the Gauss-Newton routine. In
Fig.~\ref{rcresults} we compare the VA prediction for the vortex-ring's
radius, $r_{c}$, to the numerically found values for $m=5$.
%
%
We observe that the radii provided by both methods are virtually identical,
allowing one to use the VA-predicted radius in applications, that may be
useful in experimental situations.

\begin{figure}[tbh]
\begin{center}
\includegraphics[width=3.25in]{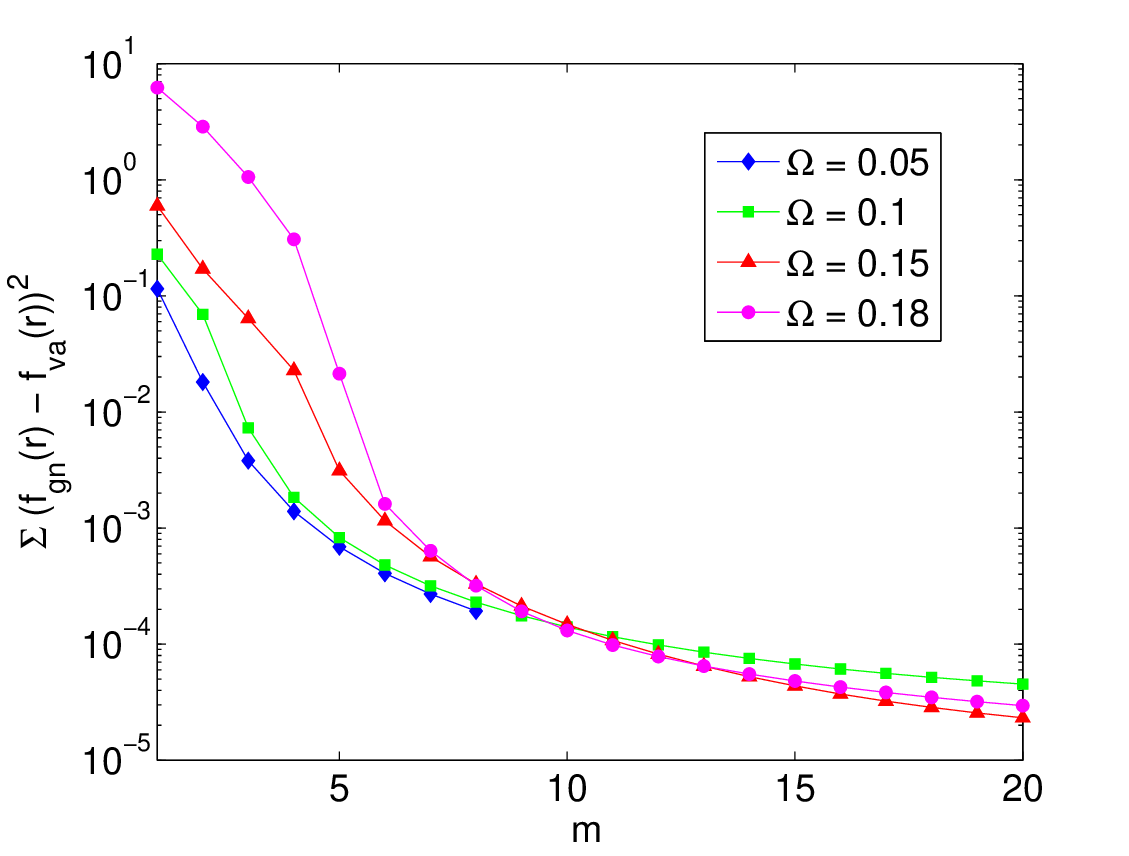} %
\includegraphics[width=3.25in]{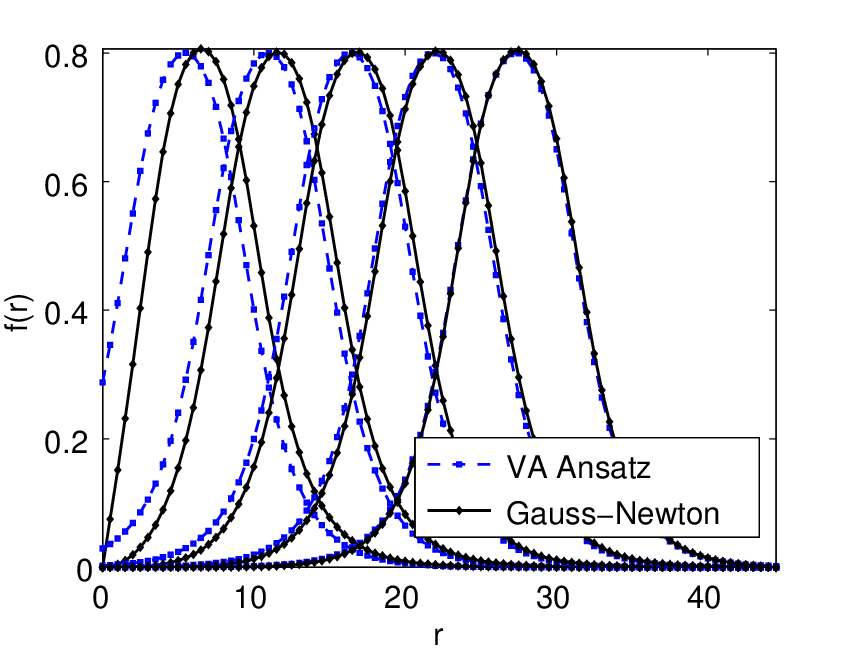}
\end{center}
\vspace{-0.3cm}
\caption{Top: The sum of squared errors between the variationally predicted
and numerically found radial profiles of the vortex soliton. 
Bottom:
Examples of the variational (dashed) and numerical (solid) profiles for $%
\Omega =0.15$. Shown from left to right are the profiles for $m=1,...,5$. It
is seen that the error decreases as $m$ increases. The series of error
values for $\Omega =0.05$ and $0.1$ terminate at $m=8$ because at that point
the variational profiles are already within the prescribed tolerance of $%
10^{-6}$ of the Gauss-Newton routine. }
\label{VA_GN_SS}
\end{figure}

To get an even better idea of how close the variational approach is to the
Gauss-Newton profile overall, in Fig.~\ref{VA_GN_SS} we compare the relative
sum of squared deviations of the variational profiles from their the
numerically found counterparts, for different values of $m$ and different
values of $\Omega $. We also plot a series of profiles produced by the VA
and by the Gauss-Newton method for $\Omega =0.15$. We observe that, as $m$
increases, the mismatch between the variational and numerical profiles
decreases. This is understandable, as we have neglected the $1/r_{c}$ term
in the Laplacian, while formulating the variational ansatz. Therefore, since
vortices with smaller $m$ have smaller inner radii, the discrepancy between
the variational ansatz and the numerically exact solution are expected for
lower values of $m$. The total discrepancy is observed to be quite low
overall, showing that the variational profile provides for a very accurate
rendition of the true solution, especially for larger values of $m$. Since
our VA provides an accurate radial profile of the vortex solitons, we can use
it to derive fully analytic azimuthal modulational stability predictions, which is done below.

\section{The Azimuthal Modulational Stability: Analytical Results}

\label{sec:azim}

With the profiles of the steady-state vortices available, we proceed to the
study of their azimuthal modulational stability. To this end, we apply the
methodology utilized in previous works, dealing with circular gap solitons
in the model of the circular Bragg grating \cite{Koby}, and in the cubic NLS
equation \cite{NLSxMI}: we will derive an azimuthal equation of motion by
assuming a frozen radial profile, and then perform a perturbation analysis
to analyze the stability of azimuthal perturbation modes.

In Ref.~\cite{NLSxMI}, numerically computed eigenmodes of perturbations
around the solitary vortices featured a slight coupling in the radial and
azimuthal directions, hence assuming the compete separability of the radial
and azimuthal directions for the evolution of the perturbations may lead to
discrepancies between the predictions for the stability and
numerical results. Nonetheless, the coupling between the radial and
azimuthal directions is attenuated with the increase of $m$, hence, the
predictions made under the assumption of the separability should be more
accurate for higher values of $m$. The trend to the improvement in the
accuracy of the analytical approximation with the increase of $m$ was
observed in the numerical results presented in Sec.~\ref{sec:NOresults}.

\subsection{The Azimuthal Equation of Motion}

\label{sec:azeq}

To derive the azimuthal equation of motion, we first insert the separable
ansatz (\ref{var_sep}) into the Lagrangian density (\ref{Ldence}) to obtain
\begin{eqnarray}
\mathcal{L}&=&\frac{i}{2}f^{2}(r)\,(AA_{t}^{\ast }-A^{\ast }A_{t})+\left(
\frac{df}{dr}\right) ^{2}\left\vert A\right\vert ^{2}+
\\[1.0ex]
&&
\frac{1}{r^{2}}%
f^{2}(r)\left\vert A_{\theta }\right\vert ^{2}-\frac{1}{2}%
f^{4}(r)\,\,|A|^{4}+\frac{1}{3}f^{6}(r)\,\,|A|^{6}.
\notag
\end{eqnarray}
Using our steady-state radial profiles, we can perform the integration of
the Lagrangian density over $dr$, thus arriving at an effectively
one-dimensional (in the direction of $\theta $) Lagrangian that may be used
to derive the equation of motion for $A(\theta ,t)$:
\begin{equation}
L_{\mbox{\scriptsize 1D}}=\int_{0}^{2\pi }\!\mathcal{L}_{%
\mbox{\scriptsize
1D}}\,d\theta ,
\end{equation}
\begin{eqnarray}
\mathcal{L}_{\mbox{\scriptsize 1D}}&\equiv& \frac{i}{2}C_{1}(AA_{t}^{\ast
}-A^{\ast }A_{t})+C_{2}|A|^{2}
\label{L1D}
\\[1.0ex]
&&
\notag
+C_{3}|A_{\theta }|^{2} 
-\frac{s_{1}}{2} C_{4}|A|^{4}-\frac{s_{2}}{3}C_{5}|A|^{6}, 
\end{eqnarray}
and $C_{j}$, $j=1,...,5$, are the radial integrals defined as per Eq.~(\ref%
{constants}). Evaluating the variational derivative of the action functional
\cite{VA}, and applying a linear transformation,
\begin{equation}
A\rightarrow A\,\exp (-iC_{2}t/C_{1}),~dt\rightarrow C_{3}t/C_{1},
\label{rescale}
\end{equation}
yields the following evolution equation for $A(\theta ,t)$:
\begin{equation}
iA_{t}=-A_{\theta \theta }-\left( C_{4}/C_{3}\right) |A|^{2}A+\left(
C_{5}/C_{3}\right) |A|^{4}A.  \label{Aeq}
\end{equation}%
This equation provides a description of how the radially-frozen, azimuthally
time-dependent solution will evolve in the CQNLS model.

\subsection{The Stability Analysis}

\label{sec:stable}

To study the azimuthal modulational stability of vortex-soliton
solutions to the CQNLS equation, we performed a perturbation
analysis in the framework of the azimuthal equation of motion
(\ref{Aeq}), with the objective to compute the growth rates of small
perturbations. We begin with an azimuthal ``plane-wave" solution
perturbed by a complex time-dependent perturbation:
\begin{equation}
A(\theta ,t)=\left[ 1+u(\theta ,t)+iv(\theta ,t)\right] \,e^{i(m\theta
+\Omega ^{^{\prime }}t)},  \label{Apert}
\end{equation}%
with $|u|,|v|\ll 1$. Inserting this into Eq.~(\ref{Aeq}), performing the
linearization with respect to the perturbations, and separating the result
into real and imaginary parts, we obtain a system of coupled equations (a
time shift was made here, $t\rightarrow t+\,\theta /\left( 2m\right) $):
\begin{equation}
\begin{array}{ll}
u_{t}=-v_{\theta \theta }, &  \\[2ex]
v_{t}=u_{\theta \theta }+\left( \dfrac{2C_{4}-4C_{5}}{C_{3}}\right) \,u. &
\end{array}
\label{uvPDE}
\end{equation}

In order to study the azimuthal modulational stability of the solitary
vortices, we expand $u$ and $v$ into Fourier series over azimuthal harmonics
with integer wavenumbers $K$:
\begin{eqnarray}
\hat{u}(K,t)&=&\int_{0}^{2\pi }\!u(\theta ,t)\,e^{iK\theta }\,d\theta ,
\notag
\\[1.0ex]
\hat{v}(K,t)&=&\int_{0}^{2\pi }\!v(\theta ,t)\,e^{iK\theta }\,d\theta .
\label{uvtrans}
\end{eqnarray}
Applying these transforms to Eq.~(\ref{uvPDE}) yields two coupled equations
for the amplitudes of $u$ and $v$ of each mode. In a matrix form, the equations are
\begin{equation}
\frac{d}{dt}\left[
\begin{array}{c}
\hat{u} \\[3.5ex]
\hat{v}%
\end{array}%
\right] =\left[
\begin{array}{cc}
0 & K^{2} \\[2ex]
\left( \dfrac{2C_{4}-4C_{5}}{C_{3}}-K^{2}\right) & 0%
\end{array}%
\right] \!\left[
\begin{array}{c}
\hat{u} \\[3.5ex]
\hat{v}%
\end{array}%
\right] .  \label{muv}
\end{equation}%
The growth rates for each azimuthal wavenumber $K$ are simply eigenvalues of
Eq.~(\ref{muv}). Taking into account the underlying rescaling (\ref{rescale}%
), they are:
\begin{equation}
\lambda _{1,2}=\pm \frac{C_{3}}{C_{1}}\sqrt{K^{2}\left( K_{%
\mbox{\scriptsize
crit}}^{2}-K^{2}\right) },  \label{grates}
\end{equation}%
where $K_{\mbox{\scriptsize crit}}$ is the critical value of $K$, above
which all the modes are stable:
\begin{equation}
K_{\mbox{\scriptsize crit}}\equiv \sqrt{\dfrac{2\left( C_{4}-2C_{5}\right) }{%
C_{3}}}.  \label{kcrit}
\end{equation}

For the modulationally unstable vortices, it is useful to know the
maximum growth rate and the mode that exhibits it.  This is because, in an
experiment, even an unstable vortex may be practically ``stable enough" if the maximum perturbation growth rate is small
enough. This information can be extracted from
expression~(\ref{grates}) by equating its derivative to zero and
solving for $K$, which reveals the fastest growing perturbation mode
(and subsequently, the prediction of the number of fragments that
the unstable vortices will break up into):
\begin{equation}
K_{\mbox{\scriptsize max}}=\sqrt{\frac{C_{4}-2C_{5}}{C_{3}}}=\frac{1}{\sqrt{2%
}}K_{\mbox{\scriptsize crit}}.  \label{kmax}
\end{equation}%
We can then insert this value into Eq.~(\ref{kcrit}), which yields the
maximum growth rate,
\begin{equation}
\lambda _{\mbox{\scriptsize max}}=\frac{C_{4}-2C_{5}}{C_{1}}=\frac{C_{3}}{%
2C_{1}}K_{\mbox{\scriptsize crit}}^{2}.  \label{lmax}
\end{equation}%
In fact, because $K$ must be integer, the actual fastest growing eigenmode
may correspond to the integer $K$ closest to value (\ref{kmax}). Accordingly,
the actual largest growth rate may be somewhat smaller than the one given by
Eq.~(\ref{lmax}).

For the vortex to be azimuthally stable against all modes, one needs either $%
K_{\mbox{\scriptsize
crit}}<1$ (then, there is no integer value $K<K_{\mbox{\scriptsize
crit}}$) or $K_{\mbox{\scriptsize
crit}}$ being purely imaginary. According to Eq.~(\ref{kcrit}), these two
stability criteria amount to the following inequalities:
\begin{equation}
\mathrm{imaginary}~~K_{\mbox{\scriptsize crit}}:~~C_{4}-2C_{5}<0;
\label{Kcriteria1}
\end{equation}%
\begin{equation}
\left\vert K_{\mbox{\scriptsize crit}}\right\vert
<1:~~C_{4}-2C_{5}-C_{3}/2<0.  \label{Kcriteria2}
\end{equation}%
Since $C_{3}$ is positive, the former condition is stricter than the latter
one. Although the latter condition is sufficient for the azimuthal
stability, we also keep the former one (according to the above derivation,
it is relevant to the infinite line) because it leads to an estimation for
the critical frequency which is independent of charge $m$ (see below).

For our predictions, we compute the $C_i$ constants using numerical integrals of the numerically exact radial profiles.  However, since the variational analytic profiles are so close to the numerically exact profiles, one can use expressions (\ref{cconstsva}) to find approximate analytic values of  Eqs.~(\ref{Kcriteria1})
and (\ref{Kcriteria2}), which then yield critical values $\Omega _{\mbox{\scriptsize st}}(m)$
of $\Omega $, above which all the vortex solitons are azimuthally stable for charge $m$.  These values can then be compared to those computed by the numerically exact profiles.

\subsection{Analytical Stability Predictions Using the Variational Profile}

\label{sec:miva}

As we showed in Sec.~\ref{sec:NOresults}, the variational ansatz yields a
useful approximation to the true radial profiles. Therefore, the VA can be
employed to calculate the constants in Eq.~(\ref{cconstsva}), and thereby
derive analytical predictions for the azimuthal modulational stability. In
Sec.~\ref{sec:stable} it was demonstrated that, for studying the stability
of the vortices, only the values of $K_{\mbox{\scriptsize crit}}$ and $%
C_{3}/C_{1}$ are required,
\begin{equation}
K_{\mbox{\scriptsize crit}}^{\mbox{\scriptsize va}}=m\sqrt{\frac{6\Omega
^{\ast }-\left( 15/8\right) +5\sqrt{3\Omega ^{\ast }}/\left( 4T\right) }{%
\Omega ^{\ast }-G(\Omega ^{\ast })}},
\end{equation}
\begin{equation}
\left( \frac{C_{3}}{C_{1}}\right) _{\mbox{\scriptsize va}}\!=\frac{\Omega
^{\ast }-G(\Omega ^{\ast })}{m^{2}},
\end{equation}%
since the other coefficients can be expressed in terms of these two (here,
as before, $\Omega =G(\Omega ^{\ast})$).

\begin{figure}[tbh]
\begin{center}
\includegraphics[width=3.25in]{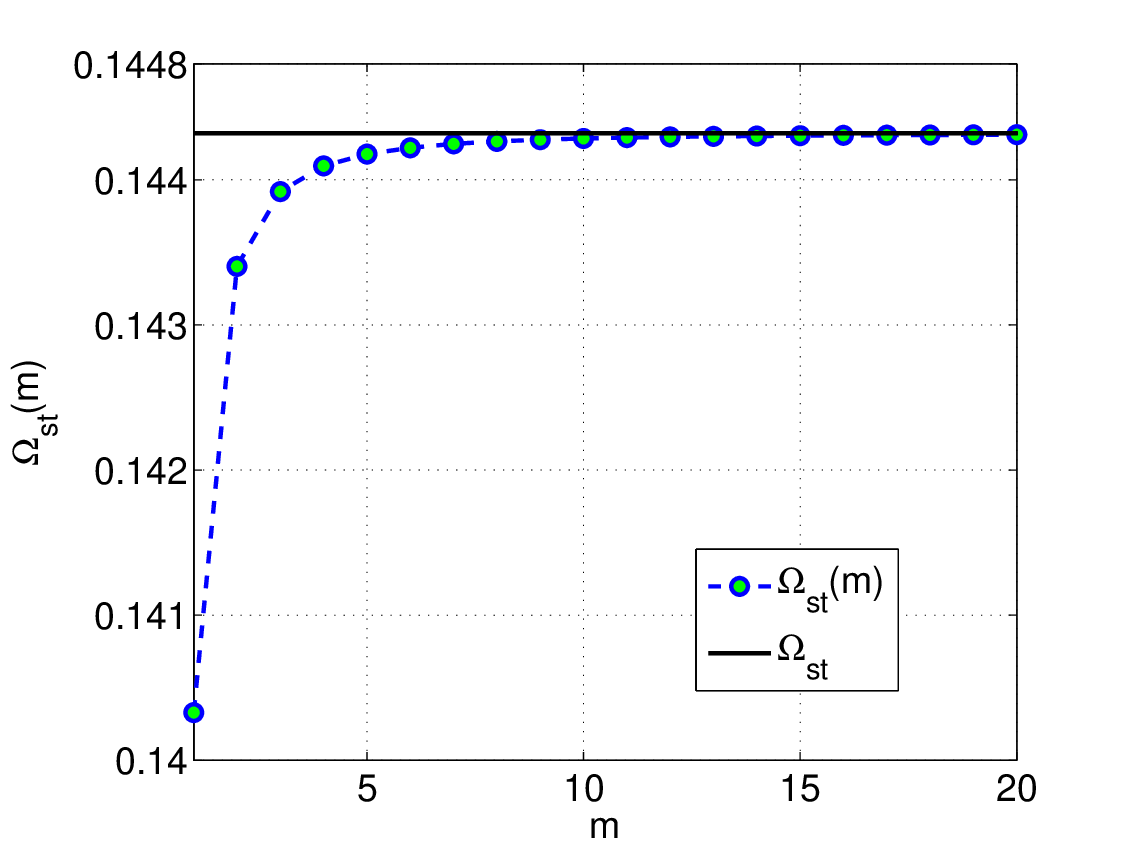} %
\includegraphics[width=3.25in]{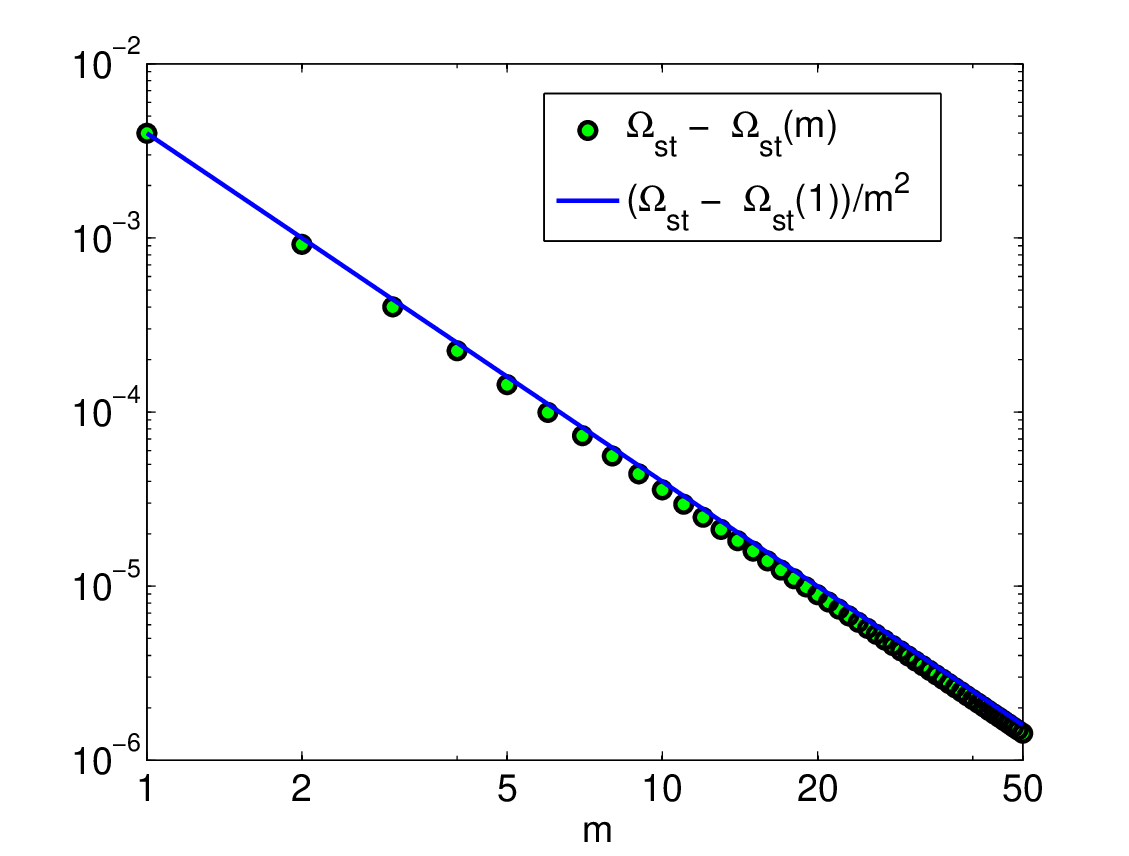}
\end{center}
\vspace{-0.3cm}
\caption{(Color online) 
Top: The critical value of frequency
$\Omega_{\mbox{\scriptsize st}}^{\mbox{\scriptsize va}}(m)$ versus 
topological charge $m$, according to the prediction of the variational approach. 
Bottom: The rate of the convergence between 
$\Omega _{\mbox{\scriptsize st}}^{\mbox{\scriptsize va}}(m)$ and 
$\Omega _{\mbox{\scriptsize st}}^{\mbox{\scriptsize va}}$ at 
increasing $m$. The plotted curve,
$\mathrm{const}\times m^{-2}$, which starts from our $m=1$ computed value, 
demonstrates that
the convergence rate is proportional to $1/m^{2}$.}
\label{omstbvm}
\end{figure}

The VA can predict the critical value of $\Omega ^{\ast }$ above which the
vortices are modulationally stable. Inserting the constants defined in Eq.~(%
\ref{cconstsva}) into the stability criteria (\ref{Kcriteria1}) and (\ref%
{Kcriteria2}), we arrive at the following expressions which determine the
critical frequency:
\begin{equation}
K_{\mbox{\scriptsize crit}}=0:~~6\Omega ^{\ast }-\frac{15}{8}+\frac{5\sqrt{%
3\Omega ^{\ast }}}{4T}=0;  \label{Kstb_eq}
\end{equation}%
\begin{equation}
K_{\mbox{\scriptsize crit}}=1:~~\sqrt{\frac{6\Omega ^{\ast }-15/8+5\sqrt{%
3\Omega ^{\ast }}/\left( 4T\right) }{\Omega ^{\ast }-G(\Omega ^{\ast })}}-%
\frac{1}{m}=0,  \label{Kstb_eq2}
\end{equation}%
where $T$ is defined as per Eq.~(\ref{T}). We see that, at $m\rightarrow
\infty $, these two criteria tend to coincide. At finite $m$, Eqs.~(\ref%
{Kstb_eq}) and (\ref{Kstb_eq2}) give different critical values of $\Omega
^{\ast }$, and hence of $\Omega $ too ---one which depends on charge $m$,
and the other one, independent of $m$, which represents an upper bound on $%
K_{\mbox{\scriptsize crit}}$ for all values of $m$. We denote the
charge-dependent critical value as $\Omega _{\mbox{\scriptsize st}}(m)$, and
the charge-independent one as $\Omega _{\mbox{\scriptsize st}}$.

Solving Eq.~(\ref{Kstb_eq}) for $\Omega ^{\ast }$ by means of a root finder
and inserting the result into Eq.~(\ref{omfunction}) yields
\begin{equation}
\Omega _{\mbox{\scriptsize st}}^{\mbox{\scriptsize va}}=0.144320424.
\label{omstbva}
\end{equation}%
Since this value is \emph{smaller} than $\Omega _{\mbox{\scriptsize max}}^{%
\mbox{\scriptsize 2D}}=3/16$, this result predicts that azimuthally stable
vortices exist for \emph{all values} of $m$, at $\Omega >\Omega _{%
\mbox{\scriptsize st}}^{\mbox{\scriptsize va}}$. We also solved Eq.~(\ref%
{Kstb_eq2}) for $\Omega $ at various values of $m$. The results are
displayed in Fig.~\ref{omstbvm}. It is seen that the critical frequency $%
\Omega _{\mbox{\scriptsize st}}^{\mbox{\scriptsize va}}(m)$ increases with
the increase of $m$ and eventually converges to $\Omega _{%
\mbox{\scriptsize
st}}^{\mbox{\scriptsize va}}$. Thus, the stability window in $\Omega $ is
larger for lower charges. In Ref.~\cite{AMIxCQNLSxNEW}, it was also
concluded that azimuthally stable vortices exist for all $m$, and that lower
charges indeed have a larger stability window. However, according to Ref.~%
\cite{AMIxCQNLSxNEW}, $\Omega _{\mbox{\scriptsize st}}=\Omega _{%
\mbox{\scriptsize
max}}^{\mbox{\scriptsize 2D}}$, i.e., the stability window shrinks towards $%
0 $ as $m\rightarrow \infty $. The estimate obtained in that work shows that
the window shrinks as $1/m^{2}$. This conclusion precludes experimental
creation of higher-order stable vortices, as the respective stability
interval would be too small, and the radius of the vortex too large.

According to the VA predictions, $\Omega _{\mbox{\scriptsize st}}$ does
increase with $m$ (and, as shown in Fig.~\ref{omstbvm}, the difference $%
\Omega_{\mbox{\scriptsize st}}-\Omega _{\mbox{\scriptsize st}}(m)$ decays
proportionally to $1/m^{2},$ which resembles the prediction of Ref. \cite%
{AMIxCQNLSxNEW}). On the other hand, since the variationally predicted value
of $\Omega _{\mbox{\scriptsize st}}$ is smaller than $3/16$, there remains a
stability window at all $m$, which does not vanish at $m\rightarrow \infty $%
.

As shown in Table~\ref{t:results}, the variational predictions are very close to those using the numerically-exact profiles.  We then note that the numerical results reported in Ref.~\cite{AMIxCQNLSxNEW} are more accurate than our predictions, the stability window indeed shrinking to zero at high values of $m$.

\begin{figure*}[tbh]
\begin{center}
\includegraphics[width=6.25in]{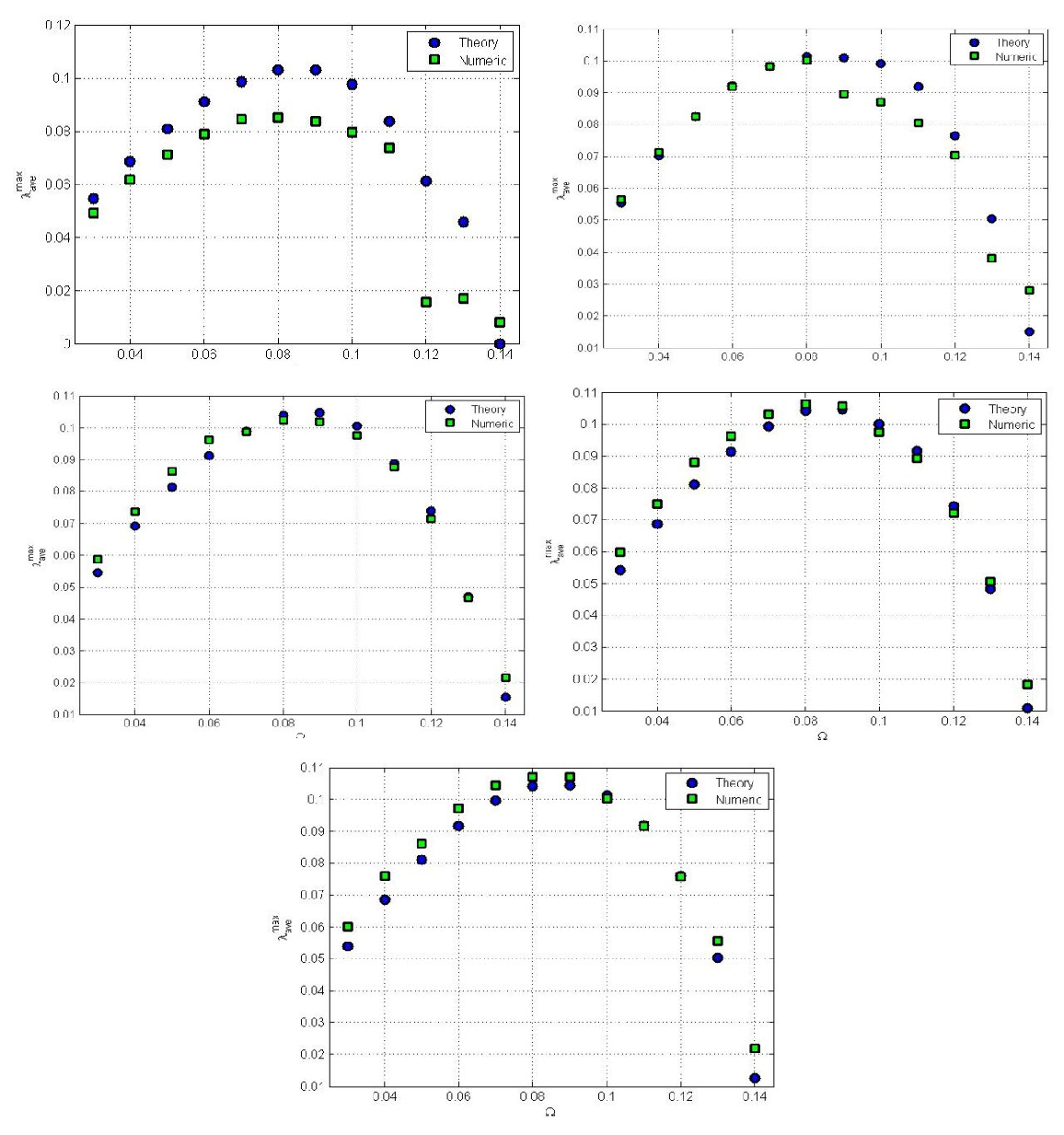}
\end{center}
\vspace{-0.3cm}
\caption{(Color online) Numerical predictions and numerical results for the
growth rates of the fastest growing unstable perturbation mode around vortex
solitons with charges $m=1,...,5$ (left to right, top to bottom). In the
interval of $\Omega \in \left[ 0.03,0.14\right] $, with the step of $\Delta
\Omega =0.01$, the growth-rate predictions (shown by circles) were obtained
by computing the integrals in the underlying Lagrangian, using the
numerically exact profiles obtained through the Gauss-Newton optimization.
We then ran full simulations of the vortex and recorded the average
instability growth rate (shown by squares). The radial, angular, and
temporal variables were discretized with steps $\Delta r=1$, $\Delta \protect%
\theta =(2\protect\pi )/(20\,\mbox{max}[m,K_{\mbox{\scriptsize max}},2])$,
and $\Delta t=0.001$, respectively. Overall, the predictions match the
numerical results very well.}
\label{AMI_RESULTS}
\end{figure*}

\section{Azimuthal Modulational Stability: Numerical Results}

\label{sec:AMSresults}

In this section we report the numerical predictions and full simulation results for the azimuthal modulational (in)stability of the vortices. For the 2D
simulations, we used a finite-difference scheme with a central
difference in space and fourth-order Runge-Kutta in time \cite{NUMODE}. We
used both polar-coordinate and Cartesian grids. The polar grid makes
computing the growth of individual perturbation modes easier, therefore we
used it to test our predictions for azimuthally unstable vortices. However,
the polar grid forces one to use smaller time steps in the finite-difference
scheme than is needed for the equivalent Cartesian grid. Therefore, for
testing the critical frequency $\Omega _{\mbox{\scriptsize st}}(m)$, which
requires running multiple long-time simulations, we use the Cartesian grid.  We used second-order differencing for the polar-coordinate simulations, and (due to the long simulation time needed) a fourth-order differencing for the Cartesian simulations.

\begin{figure}[tbh]
\begin{center}
\includegraphics[width=3.7in]{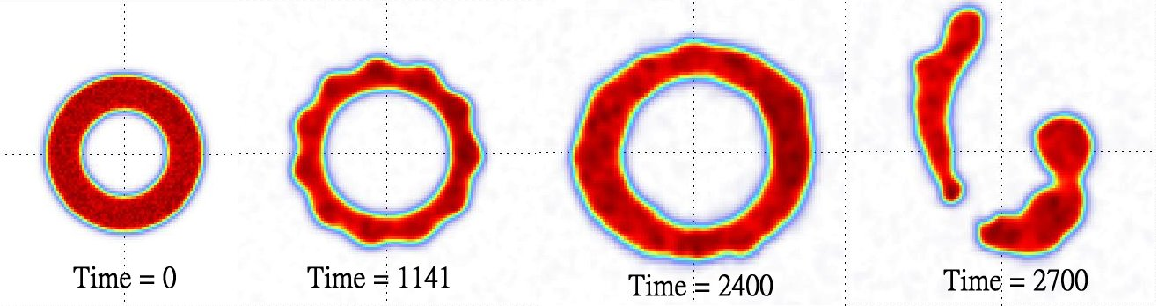}
\end{center}
\vspace{-0.3cm}
\caption{(Color online) Examples of the evolution of a vortex with charge $%
m=4$ and $\Omega =0.18$, perturbed by a random perturbation of size $\protect%
\epsilon =0.05$. Pictured is the squared absolute value of the wave
function. We have set $\Delta r=2$, $r_{\mbox{\scriptsize max}}=120$ and $%
\Delta t=0.6$. The initial vortex shape becomes deformed and then breaks up
irregularly.}
\label{snake}
\end{figure}

\subsection{Unstable Vortices}

\label{sub:results_u}

In Fig.~\ref{AMI_RESULTS} we show the results for the (in)stability of
vortices with charges $m=1,...,5$ and $\Omega \in \lbrack 0.03,0.14]$. In
general, we see a good agreement between the numerically measured growth
rates and the predictions for $m>2$; however, for values of $\Omega $ which
get closer to $3/16$, the accuracy of the predictions becomes low in each
case, implying that our predictions for $\Omega _{\mbox{\scriptsize st}}(m)$
are not precise. To further test this, we ran long-time simulations of
randomly perturbed vortices.

For $m=1,2,3$, we were easily able to identify the transition from unstable
to stable vortices, but for $m>3$ we found it very difficult to detect a
stable solution, because of a snake-like instability which breaks the vortex
into asymmetric irregular fragments. An example of this is shown in Fig.~\ref%
{snake}. Obviously, this snake-like instability is not captured by our study
of the azimuthal modulational instability. It is very plausible that it
accounts for the discrepancy between the analytic and numerical growth rates
observed in Fig.~\ref{AMI_RESULTS}.

\subsection{Stable Vortices}

\label{sec:results_s}

\begin{figure}[tbh]
\begin{center}
\includegraphics[width=2.25in]{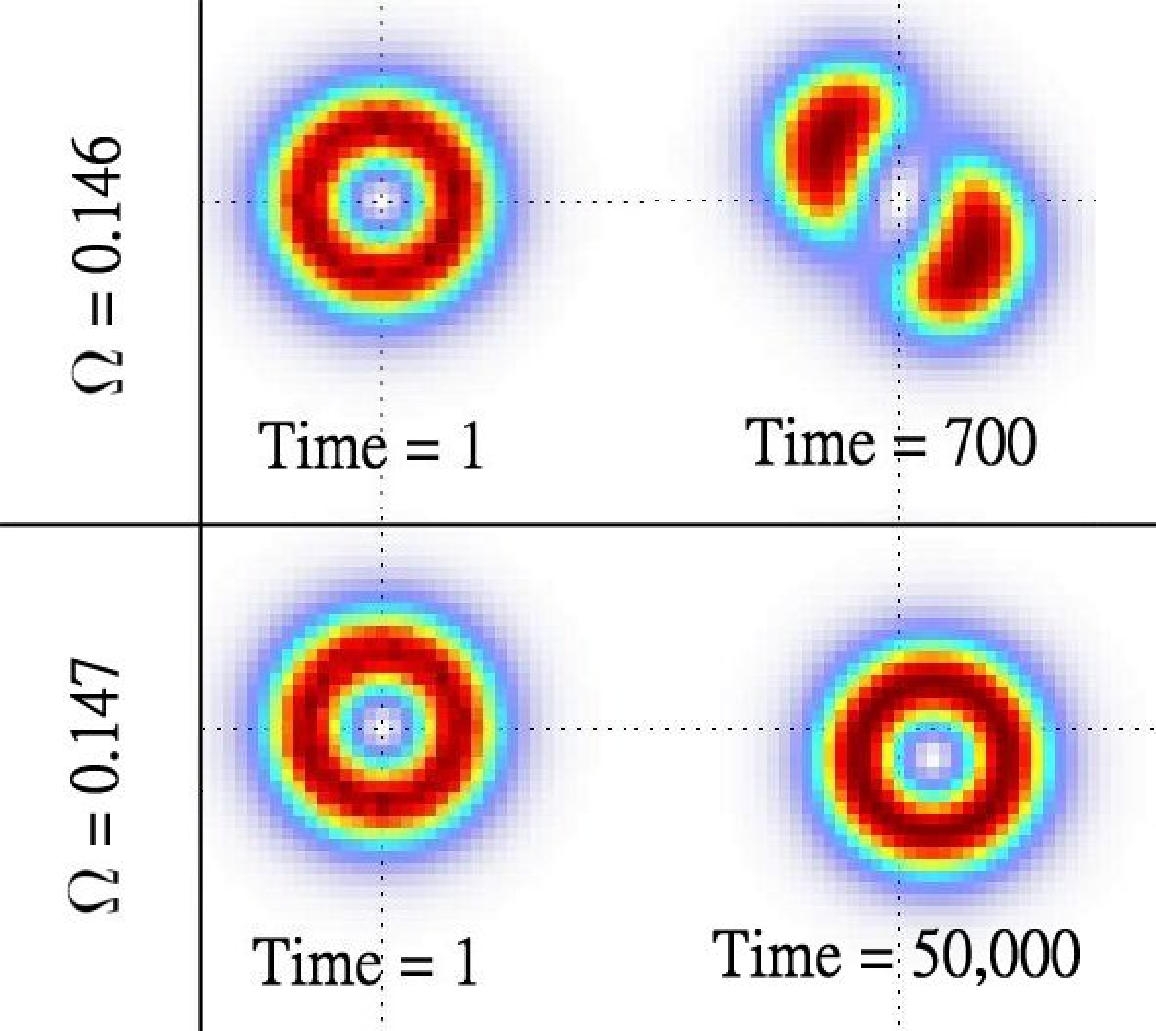}
\end{center}
\vspace{-0.3cm}
\caption{(Color online) An example of the numerical identification of the
stability border, $\Omega _{\mbox{\scriptsize st}}(m)$, for $m=1$. In this
case, the evolution of the vortex with $\Omega _{1}=0.146$ (top left) was
simulated with a random-noise perturbation in the azimuthal direction, until
it broke up into fragments due to the azimuthal instability (top right). We
then carried out long simulations of the evolution of a vortex with $\Omega
_{2}=\Omega _{1}+0.001=0.147$ (bottom left), to make it sure that the vortex
is stable ---in this example, up to $t=50,000$ (bottom right). The drift of
the position of the vortical pivot is due to the momentum imparted to it by
the random perturbations. We thus conclude that $\Omega _{
\mbox{\scriptsize
st}}(m=1)\in \left[ 0.146,0.147\right] $. In this
example, we set the grid spacing to be $\Delta x=\Delta y=1$, time step to $%
\Delta t=0.2$, and the perturbation amplitude to $\protect\epsilon =0.05$.}
\label{grmethod2}
\end{figure}

To identify the values of $\Omega _{\mbox{\scriptsize st}}(m)$ corresponding
to the stability border, we simulated the evolution of vortices with $\Omega
$ taken near the predicted values of $\Omega _{\mbox{\scriptsize st}}(m)$,
perturbed by a small uniformly distributed random noise in the azimuthal
direction. Our aim was to find a value $\Omega _{1}$ of $\Omega $ that
results in an \emph{unstable} vortex soliton (and to observe the actual
breakup), and then show that the vortex solution for $\Omega _{2}=\Omega
_{1}+0.001$ is \emph{stable}, by simulating its evolution for an extremely
long time, compared to the time necessary to reveal the full breakup in the
former case. We did this for various charges $m$. In Fig.~\ref{grmethod2},
we show the initial and final states produced by this analysis for $m=1$.

Since the snaking effect hinders our ability to simulate the evolution of
the vortices for extended time periods at $m>3$, we were not able to make
precise stability predictions in this case. However, as this effect is
dynamically distinct from the azimuthal breakup, we can still give an
approximate estimate of the critical frequency for higher charges.

In Table~\ref{t:results} we display the predictions of $\Omega _{\mbox{\scriptsize st}}(m)$ and their numerically found counterparts for $m=1,...,5$.  The column labeled ``semi-numerical" are the predictions made using the numerically exact radial profiles, while the column labeled ``VA" gives the predictions made using the variationally derived profiles.  For comparison, in the same Table we also display findings reported in earlier works. We see that most of the predictions for $m=1$ are close to the numerical result. For $m>1$, the results reported in Ref.~\cite%
{AMIxCQNLSxNEW} are most accurate ones.

\begin{table}[tbh]
\begin{center}
\begin{tabular}{|c||c||c||c|c|c|c|}
\hline
$m$ & NUM & semi-NUM & VA & Ref.~\cite{AMIxCQNLSxNEW} & Ref.~\cite%
{AMIxCQNLSxLONG} & Ref.~\cite{AMIxCQNLSxOLD} \\ \hline\hline
1 & 0.147 & 0.1399 & 0.1403 & 0.1487 & 0.16 & 0.145 \\ \hline
2 & 0.162 & 0.1433 & 0.1434 & 0.1619 & 0.17 & NC \\ \hline
3 & 0.171 & 0.1437 & 0.1439 & 0.1700 & NC & NC \\ \hline
4 & (0.178) & 0.1437 & 0.1441 & 0.1769 & NC & NC \\ \hline
5 & (0.18) & 0.1437 & 0.1442 & 0.1806 & NC & NC \\ \hline
\end{tabular}%
\end{center}
\par
\centering
\vspace{-0.3cm}
\caption{Comparison of the predictions and numerical
results for the stability border, $\Omega _{\mbox{\scriptsize
st}}(m)$.  The columns labeled ``semi-NUM" and ``VA" represent the predictions utilizing the numerically exact and the variationally derived radial profiles respectively.  The column labeled ``NUM" are the results obtained from the full 2D simulations.  Numerical findings in parentheses are
those that were hard to fix due to the emergence of the snake-like
instability (not comprised by our stability analysis), and may
therefore be less accurate than the others. We also show predictions
reported in earlier works. When no value has been computed or
reported, we
label the entry as ``NC". The predictions made in Ref.~%
\protect\cite{AMIxCQNLSxNEW} are the most accurate ones, in comparison to
our simulations.}
\label{t:results}
\end{table}

It is observed that the semi-numerical and variational results are very close, the latter ones being, in fact, slightly \emph{more accurate} in comparison with the numerical findings.  This implies that the discrepancies in the variationally derived predictions are \emph{not} due to any inherent deficiency in the VA.  Rather, the discrepancies of both sets of predictions are likely due to the approximation of the vortex soliton as a separable entity composed of radial and azimuthal parts (as discussed in Sec.~\ref%
{sec:azim}).

\begin{figure*}[tbh]
\begin{center}
\begin{tabular}{cccc}
\includegraphics[height=2.5in]{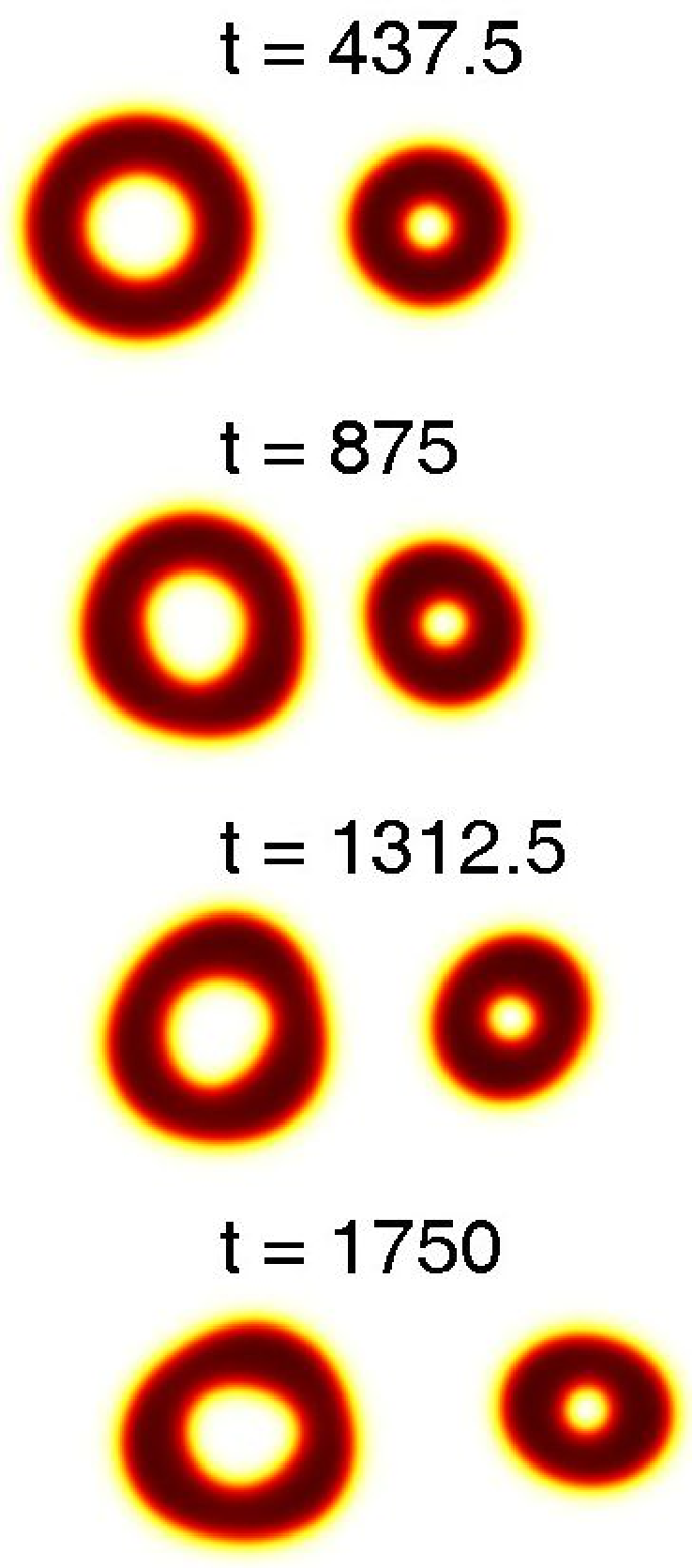}        & \hskip1.5cm %
\includegraphics[height=2.5in]{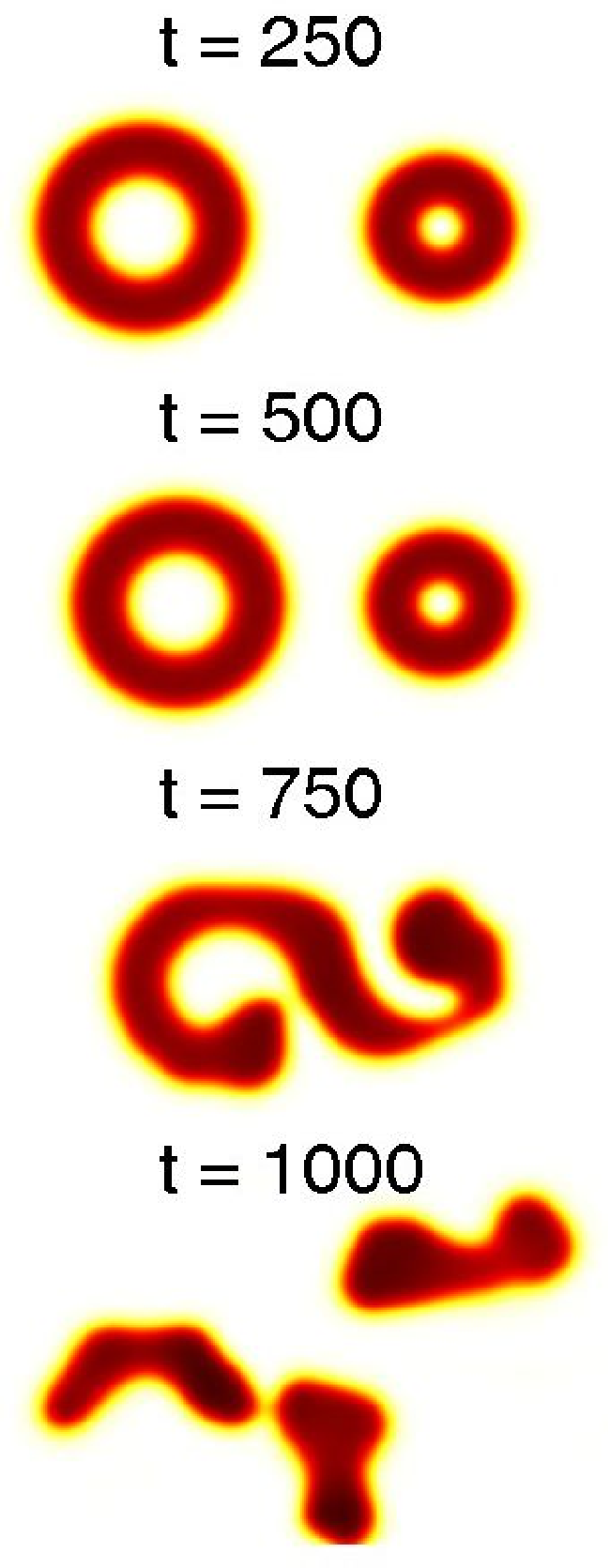}        & \hskip1.5cm %
\includegraphics[height=2.5in]{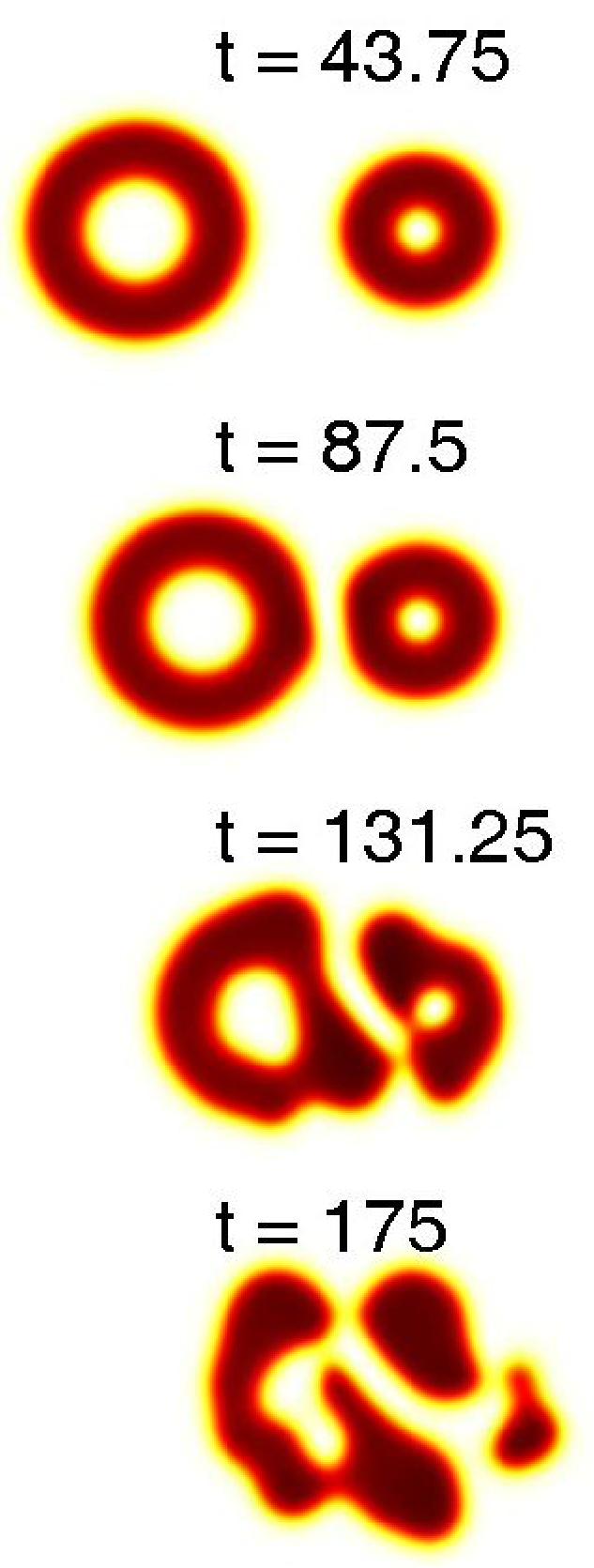} & \hskip1.5cm %
\includegraphics[height=2.5in]{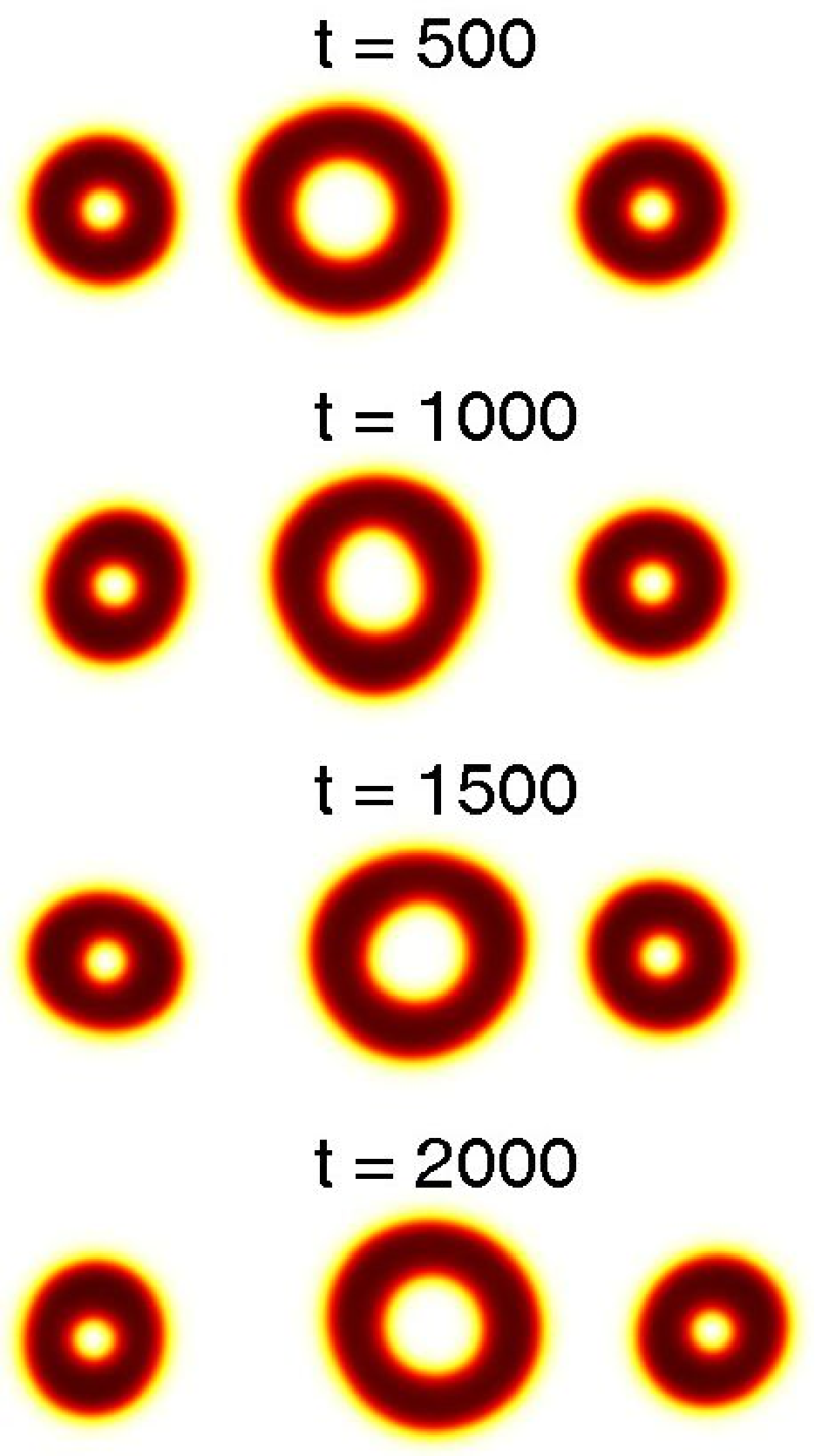}%
\end{tabular}%
\end{center}
\vspace{-0.6cm}
\caption{(Color online) Collisions between stable vortices, all with
frequency $\Omega =0.17$. First column: an elastic collision between a
moving vortex of charge $m=2$, with a slow initial velocity of $0.03$, and a
stationary vortex of charge $m=1$. The vortices have no relative phase
difference between them, which corresponds to their colliding sides being
out-of-phase, hence the interaction is repulsive. Accordingly, the vortices
undergo an elastic collision. Second column: the same as in the first
column, but with a $\protect\pi $ phase difference between the colliding
vortices, which implies that their colliding sides are in phase. This, in
turn, gives rise to a mutual attraction, which causes the vortices to merge
and eventually break up into fragments. Third column: same as in the first
column, but for a larger collisional velocity of $0.3$. The mutual repulsion
between the in-phase vortices is not enough to counterbalance the high
collisional momentum, and the vortices merge and break up into fragments.
Fourth column: a charge $m=1$ vortex with velocity $0.03$ undergoing an
elastic collision with an $m=2$ vortex, in the case of a $\protect\pi $
phase difference between them, which in turn collides with another $\protect%
\pi $-phase-shifted vortex of charge $m=1$. This phase arrangement
corresponds to vortices colliding with adjacent sides that are mutually $%
\protect\pi $-out-of-phase, and thus repel each other. These results attest
to the robustness of the vortices. }
\label{collision_series}
\end{figure*}

\section{Collisions and Scattering of Stable Vortices}

\label{sec:dyn}

In this section we study collisions and scattering of stable vortices in the
CQNLS. In Ref.~\cite{AMIxCQNLSxOLD}, such collisions of vortices of unit
charge were considered in a brief form. It was shown that both elastic and
destructive collisions could be observed depending on the phase difference, $%
\Delta \phi $, between the colliding vortices, as well as on their relative
velocity. Here we expand this study in three directions: i) we include
vortices of charge $m=2$; ii) we consider the critical velocity for elastic
collisions as a function of $\Delta \phi $; iii) we study the scattering of
vortices colliding at various values of impact parameters, by measuring the
scattering angles and speeds.

Summarizing results of numerous simulations, we have found that, at
sufficiently small collisional velocities\footnote{
A horizontal velocity $v$ is imprinted to each vortex by multiplying its
corresponding steady state by $\exp(i k x)$ with $k=v/2$. This factor of two
between $k$ and $v$ can be formally obtained by transforming CQNLS solutions
into a moving reference frame with velocity $v$ \cite{Belmonte}.}, the elastic or destructive character of the collision is solely
determined by the phase difference at the point of contact. This fact seems
to be independent of the charge of the vortices involved in the collision.
In Fig.~\ref{collision_series} we display four different cases that
illustrate the main features of head-on collisions between vortices with
different charges. In the first two columns of Fig.~\ref{collision_series}
we show the collision of vortices carrying charges $m=2$ and $m=1$, with
$\Delta \phi =0$ and $\Delta \phi =\pi$, respectively. It can be seen that, since the
vortex with $m=2$ has opposite phase on the collision side, with respect to
the vortex with $m=1$, the interaction is repulsive since it locally
emulates the interaction of two out-of-phase fundamental bright solitons
\cite{interaction,adhikari03}. Therefore, if the velocity is small
enough, the mutual repulsion determines the result of the collision. On the
other hand, when the vortex with charge $m=2$ is phase-shifted by $\pi $ (or
similarly, if the charge $m=2$ vortex were to be placed on the opposite side
of the charge $m=1$ vortex), the adjacent sides of the colliding vortices
are in phase. Therefore, the interaction is similar to that between two
in-phase bright solitons, which is \emph{attractive} \cite%
{interaction,adhikari03}. This results in the merger of the two vortices and
their eventual breakup into fragments.

\begin{figure*}[tbh]
\begin{center}
\includegraphics[width=16cm]{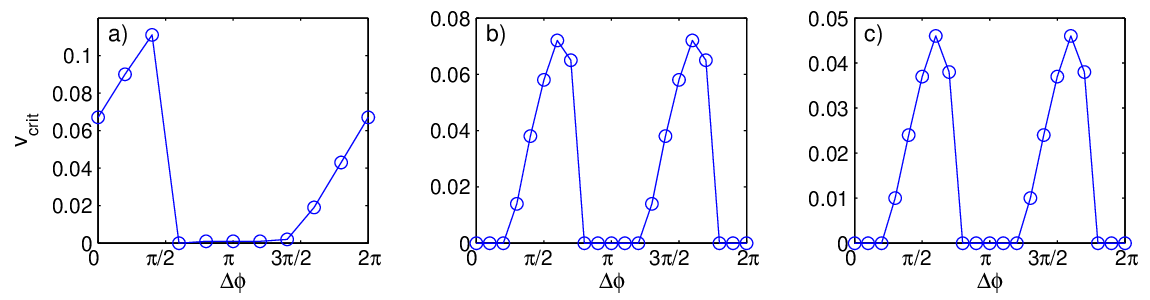}
\end{center}
\vspace{-0.6cm}
\caption{(Color online) Critical velocity $v_{\mathrm{crit}}$ for elastic
collisions between two vortices as a function of their phase difference $%
\Delta \protect\phi $. The vortex charges of the colliding vortices
correspond to: (a) $(m_{1},m_{2})=(1,1)$, (b) $(m_{1},m_{2})=(1,2)$, and (c)
$(m_{1},m_{2})=(2,2)$. All vortices are taken with intrinsic frequency $%
\Omega =0.17$. }
\label{vcrit}
\end{figure*}

\begin{figure*}[tbh]
\begin{center}
\includegraphics[width=5.4cm]{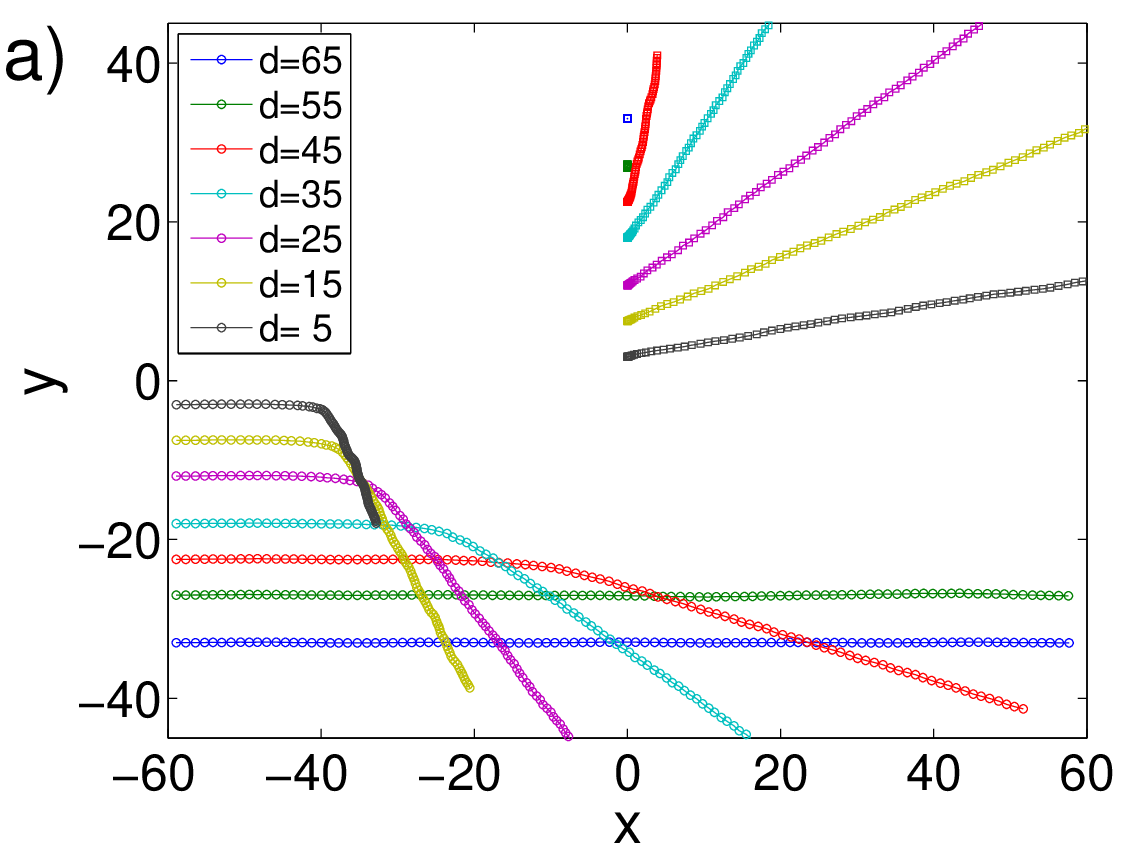} ~ %
\includegraphics[width=5.4cm]{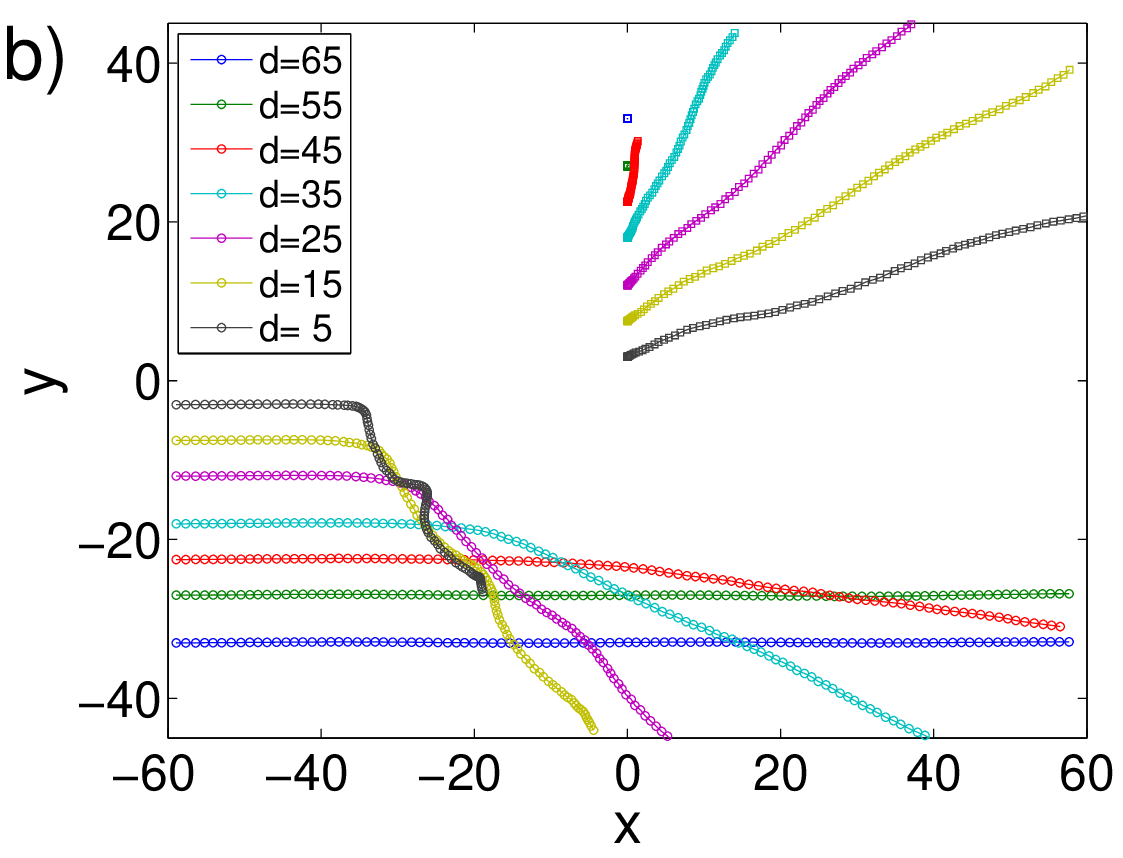}
\end{center}
\par
\begin{center}
\includegraphics[width=5.4cm]{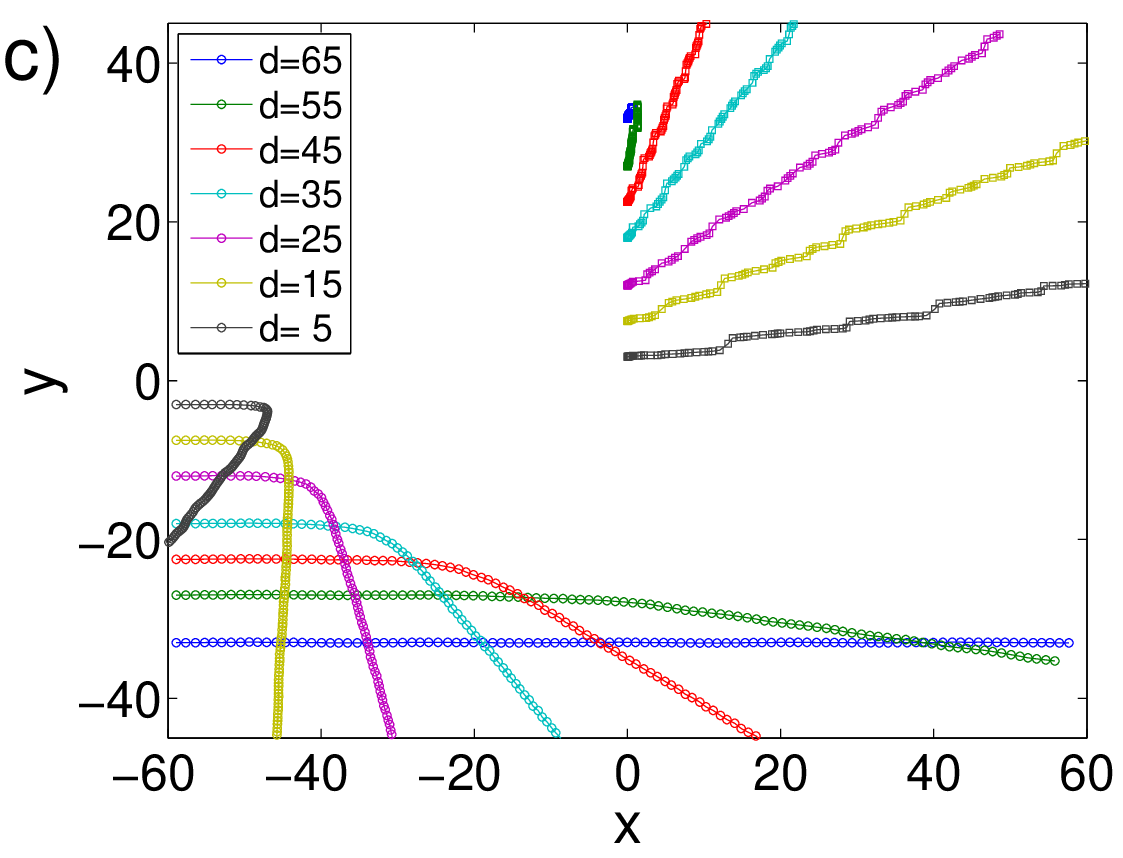} ~ %
\includegraphics[width=5.4cm]{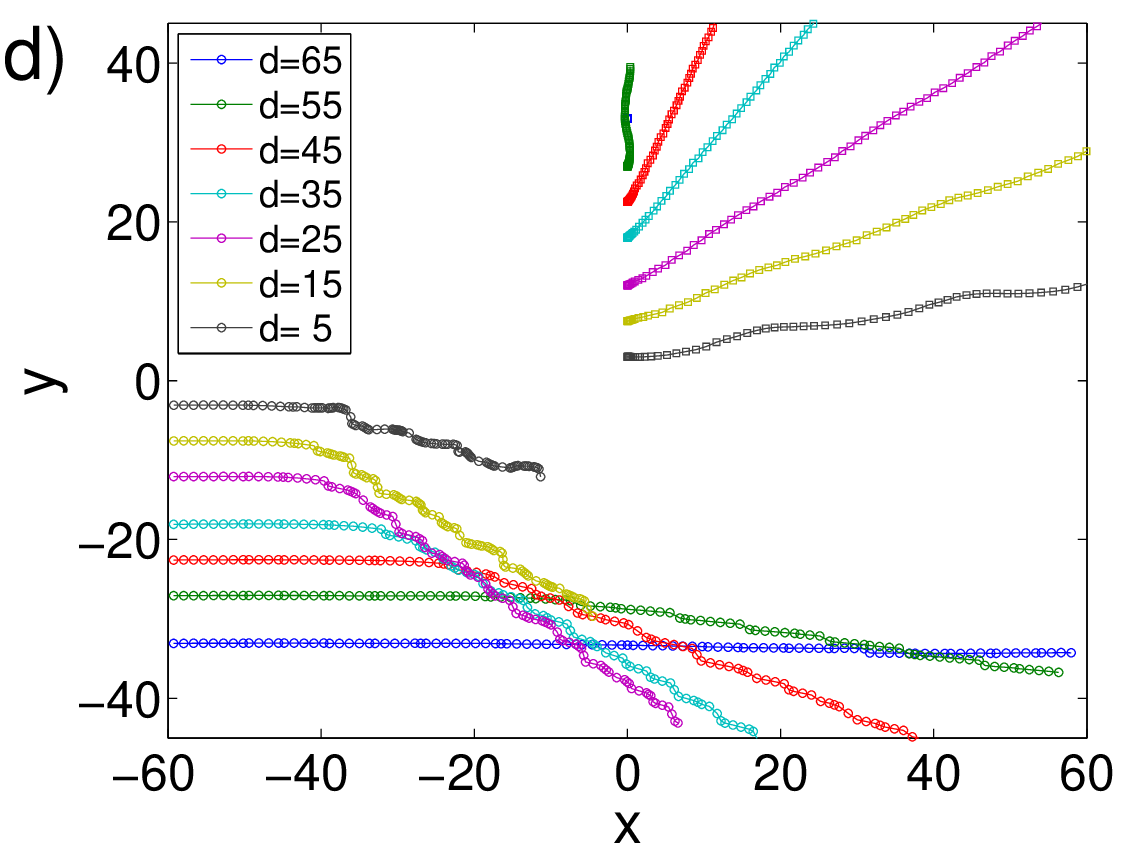}
~ \includegraphics[width=5.4cm]{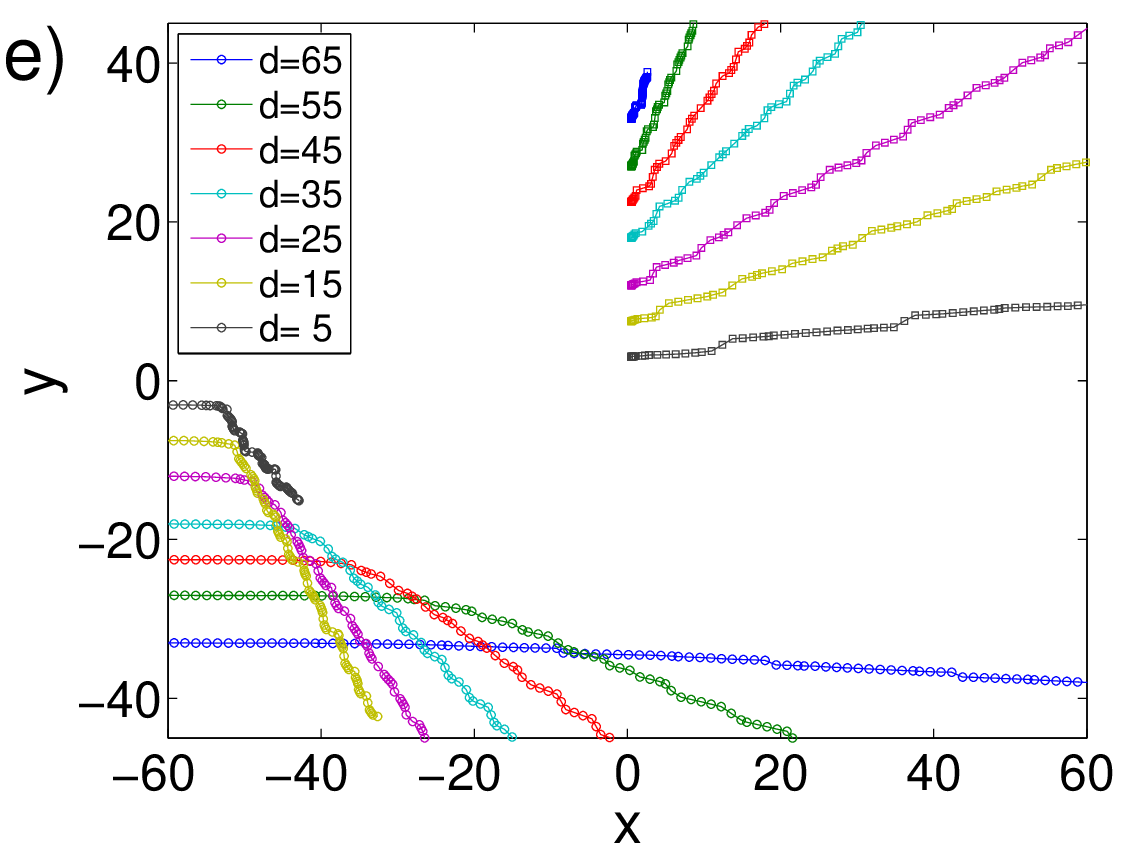}
\end{center}
\vspace{-0.6cm}
\caption{(Color online) Scattering between an incoming vortex with charge $%
m_{1}$ and velocity $v_{1}^{\mathrm{ini}}$, and a quiescent vortex of charge
$m_{2}$. The trajectory for the incoming (stationary) vortex is depicted by
small circles (squares). All panels depict different trajectories for
different values of impact parameter $d$, as indicated [$d=5,15,25,...,65$
from top to bottom (bottom to top) for the incoming (stationary) vortex].
Panels (a) and (b) correspond, respectively, to two unitary-charge vortices (%
$m_{1}=m_{2}=1$), with $\Delta \protect\phi =0$ and collision velocities $%
v_{1}^{\mathrm{ini}}=0.03$ and $v_{1}^{\mathrm{ini}}=0.06$. Panels (c) and
(d) correspond, respectively, to a $m_{1}=1$ vortex scattered by a $m_{2}=2$
one and vice versa, for the collision velocity $v_{1}^{\mathrm{ini}}=0.03$
and $\Delta \protect\phi =\protect\pi /2$. Panel (e) corresponds to a $%
m_{1}=2$ vortex scattered by a $m_{2}=2$ one, for $v_{1}^{\mathrm{ini}}=0.02$
and $\Delta \protect\phi =\protect\pi /2$. All vortices have intrinsic
frequency $\Omega =0.17$. }
\label{scattering}
\end{figure*}

\begin{figure*}[tbh]
\begin{center}
\includegraphics[width=5.4cm]{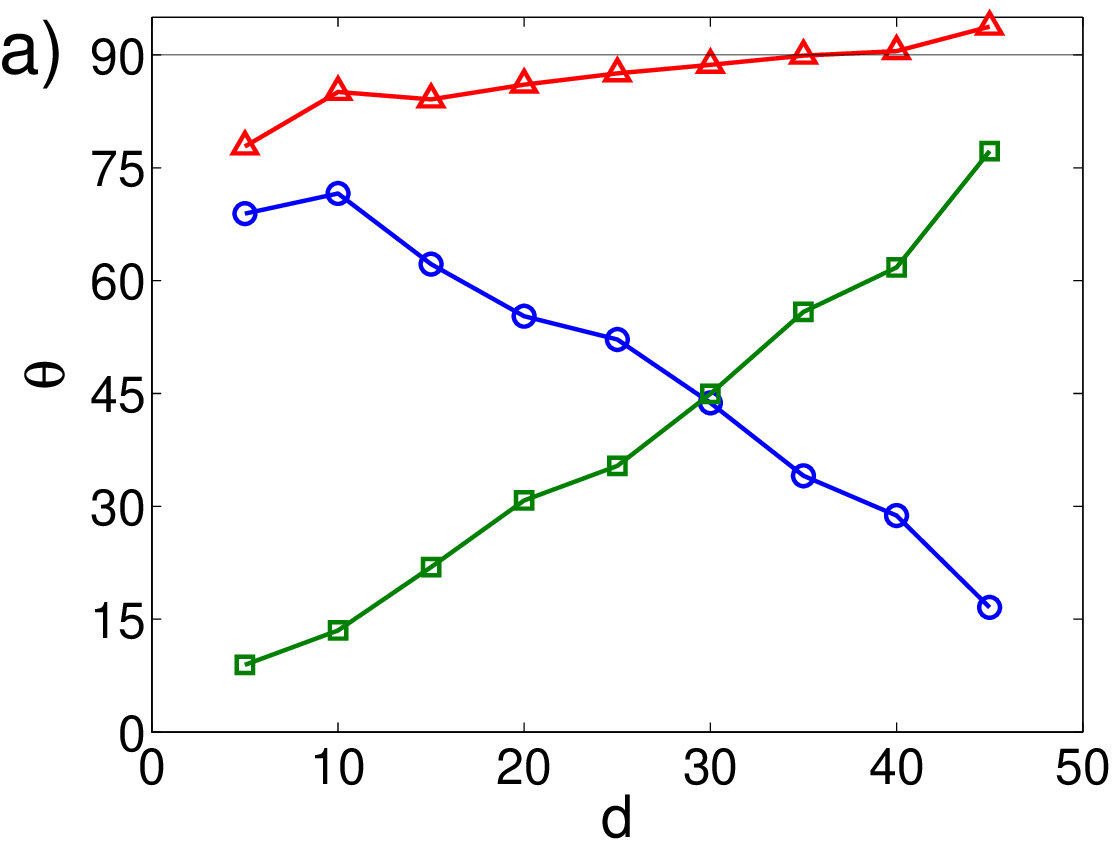} ~ %
\includegraphics[width=5.4cm]{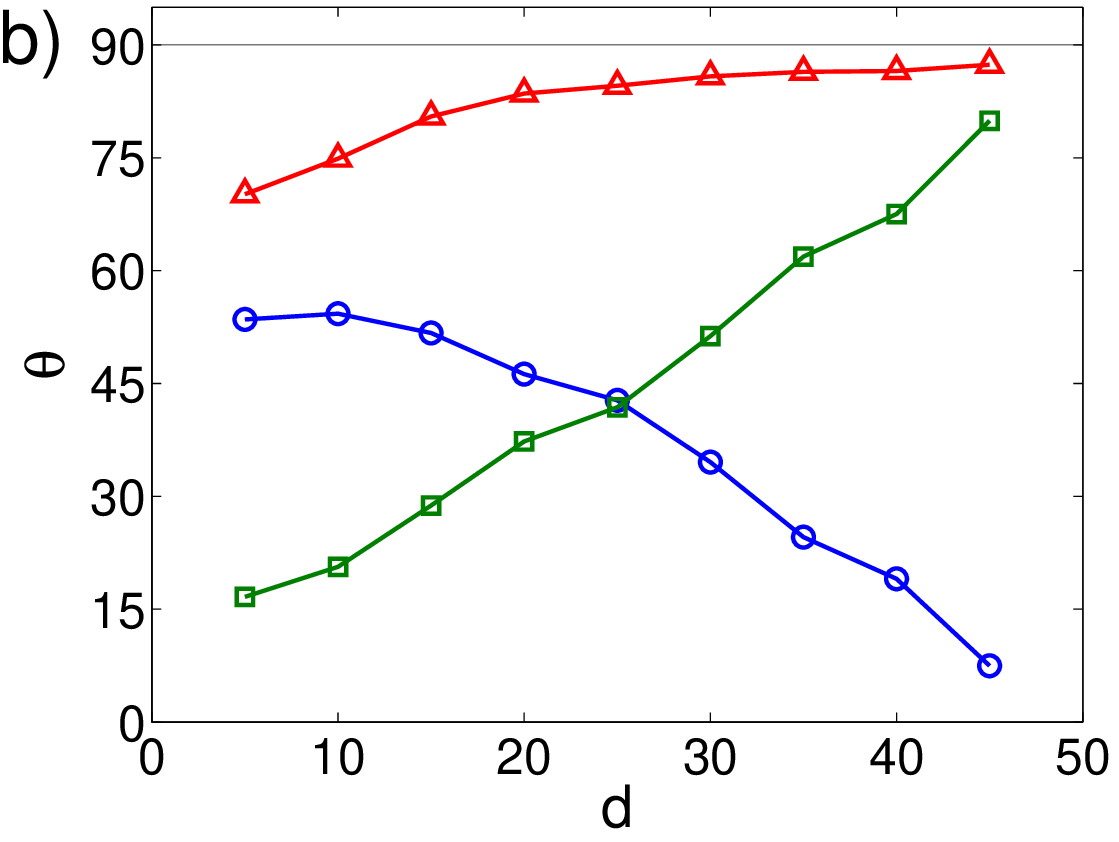}
\end{center}
\par
\begin{center}
\includegraphics[width=5.4cm]{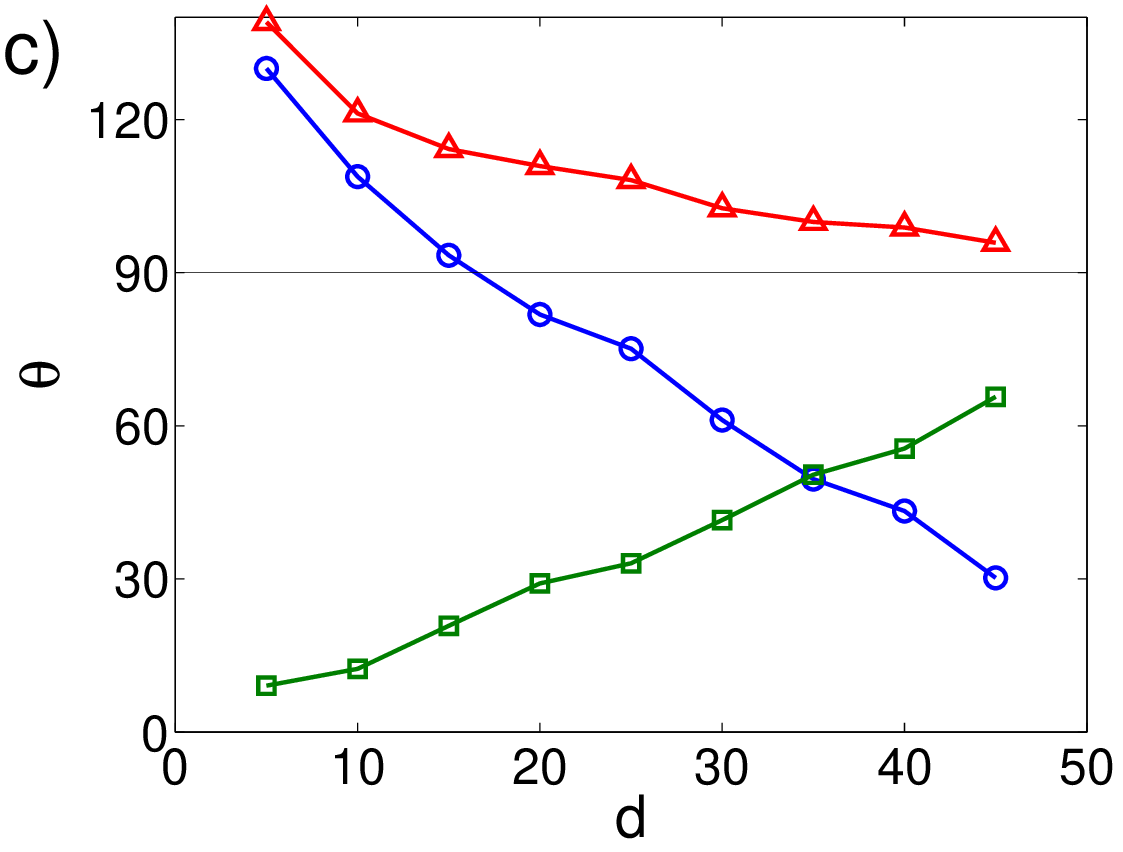} ~ %
\includegraphics[width=5.4cm]{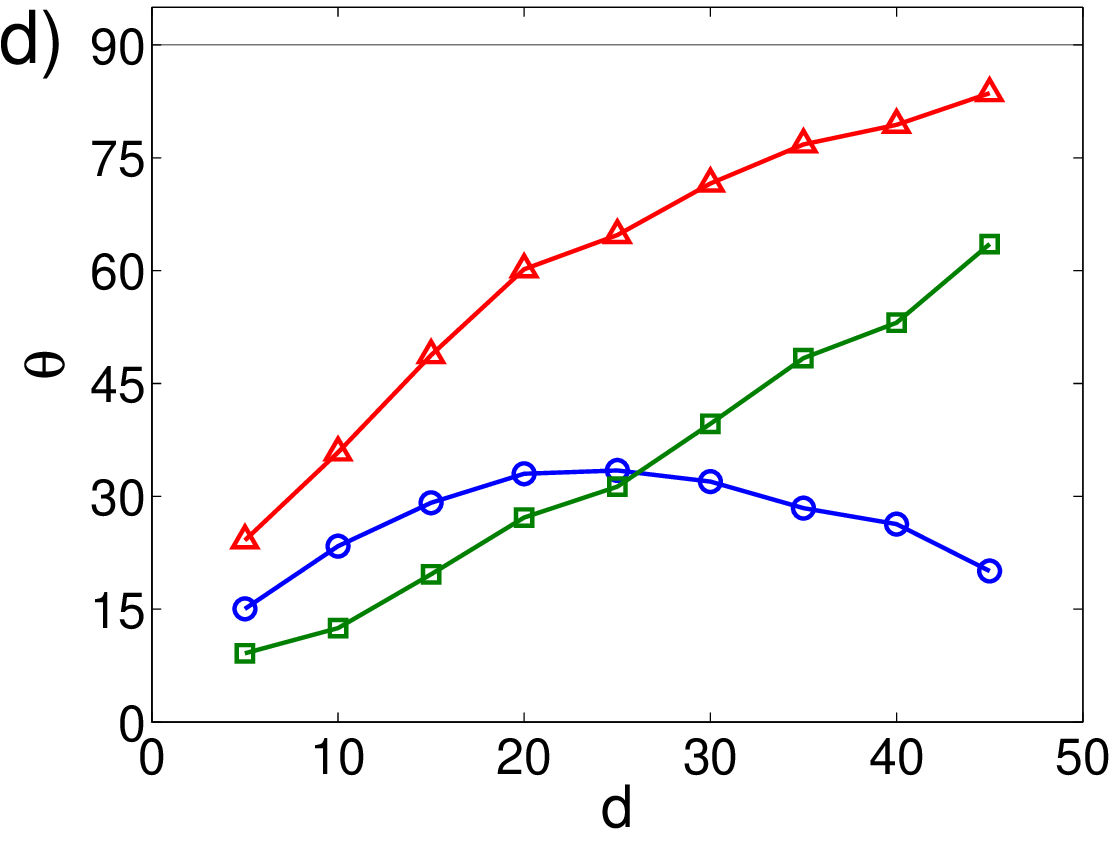}
~ \includegraphics[width=5.4cm]{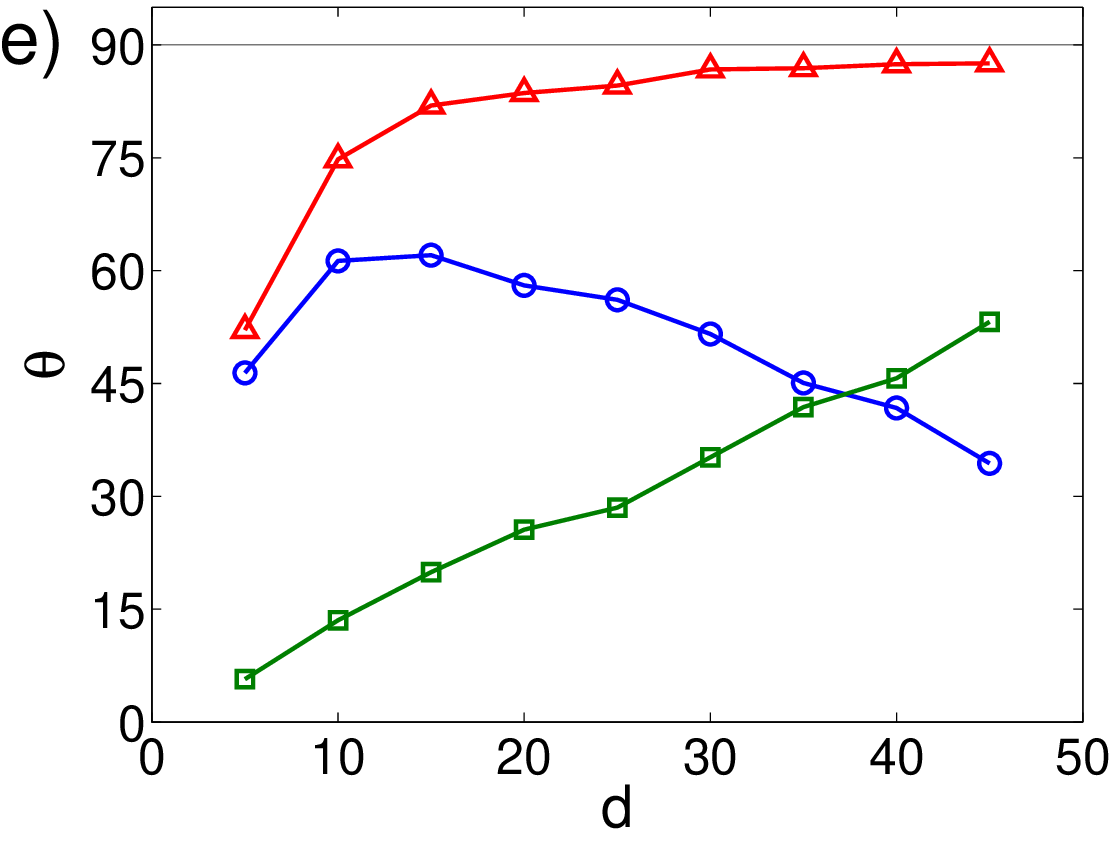}
\end{center}
\vspace{-0.6cm}
\caption{(Color online) Scattering angles for the collision of two vortices
corresponding to the panels depicted in Fig.~\protect\ref{scattering}.
Depicted by (blue) circles is the negative of the scattering angle $-\protect%
\theta _{1}$ of the incoming vortex, while (green) squares stand for the
scattering angle $\protect\theta _{2}$ of the initially quiescent vortex.
Triangles (in red) display the difference between these two angles,
$\protect \theta _{2}-\protect\theta _{1}$. All angles are given in degrees.}
\label{scat_angles}
\end{figure*}

\begin{figure*}[tbh]
\begin{center}
\includegraphics[width=5.4cm]{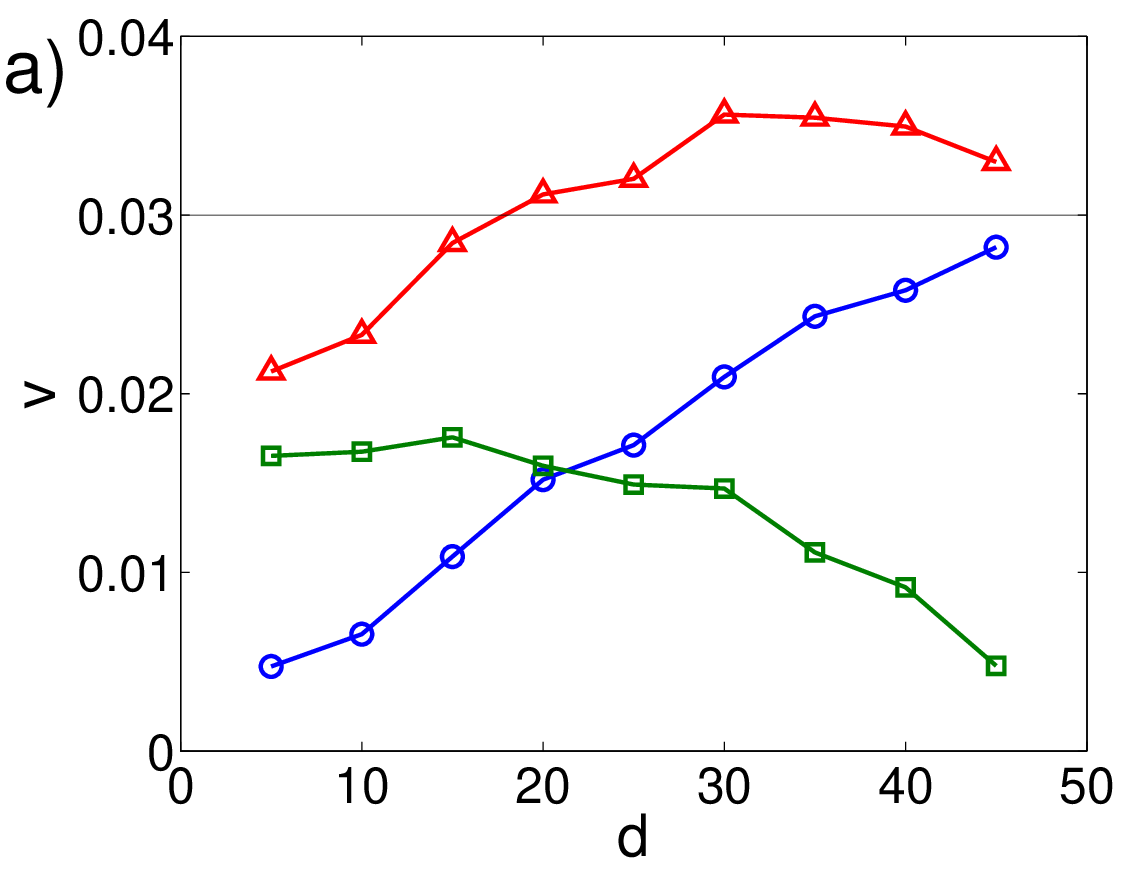} ~ %
\includegraphics[width=5.4cm]{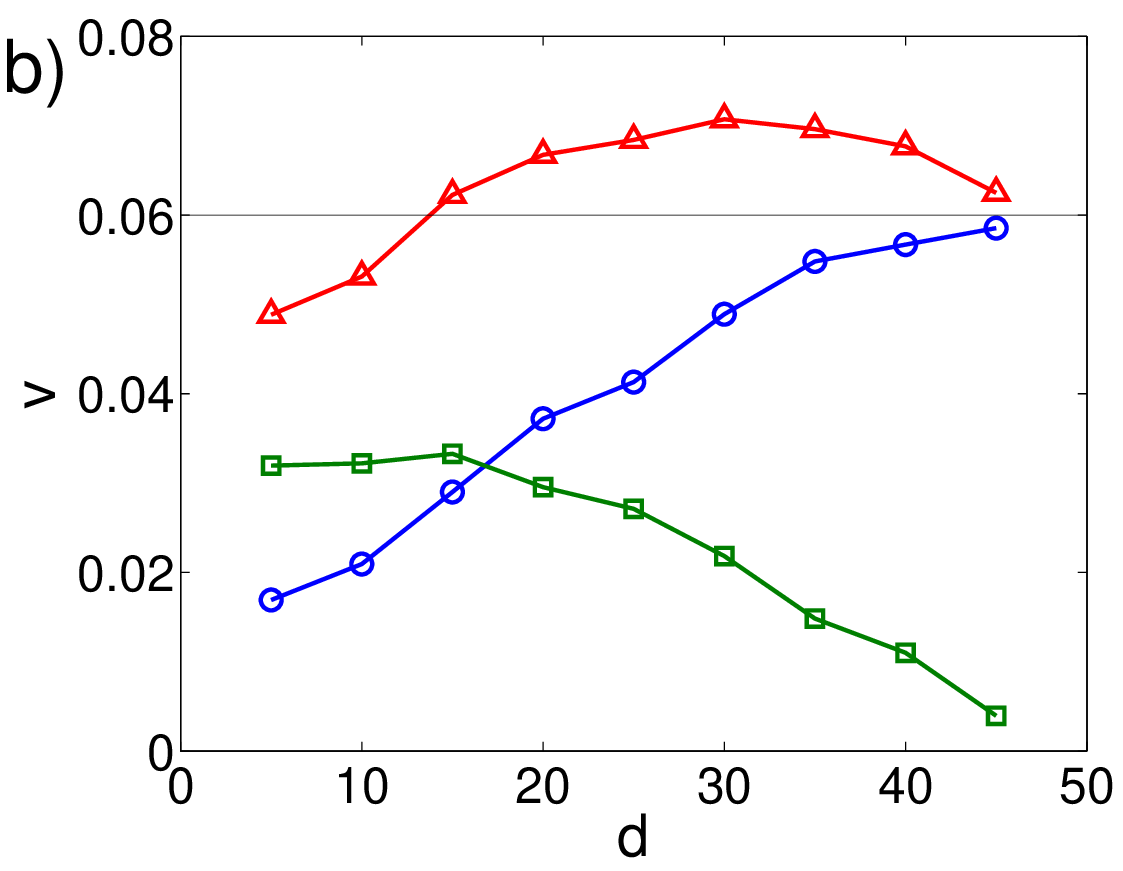}
\end{center}
\par
\begin{center}
\includegraphics[width=5.4cm]{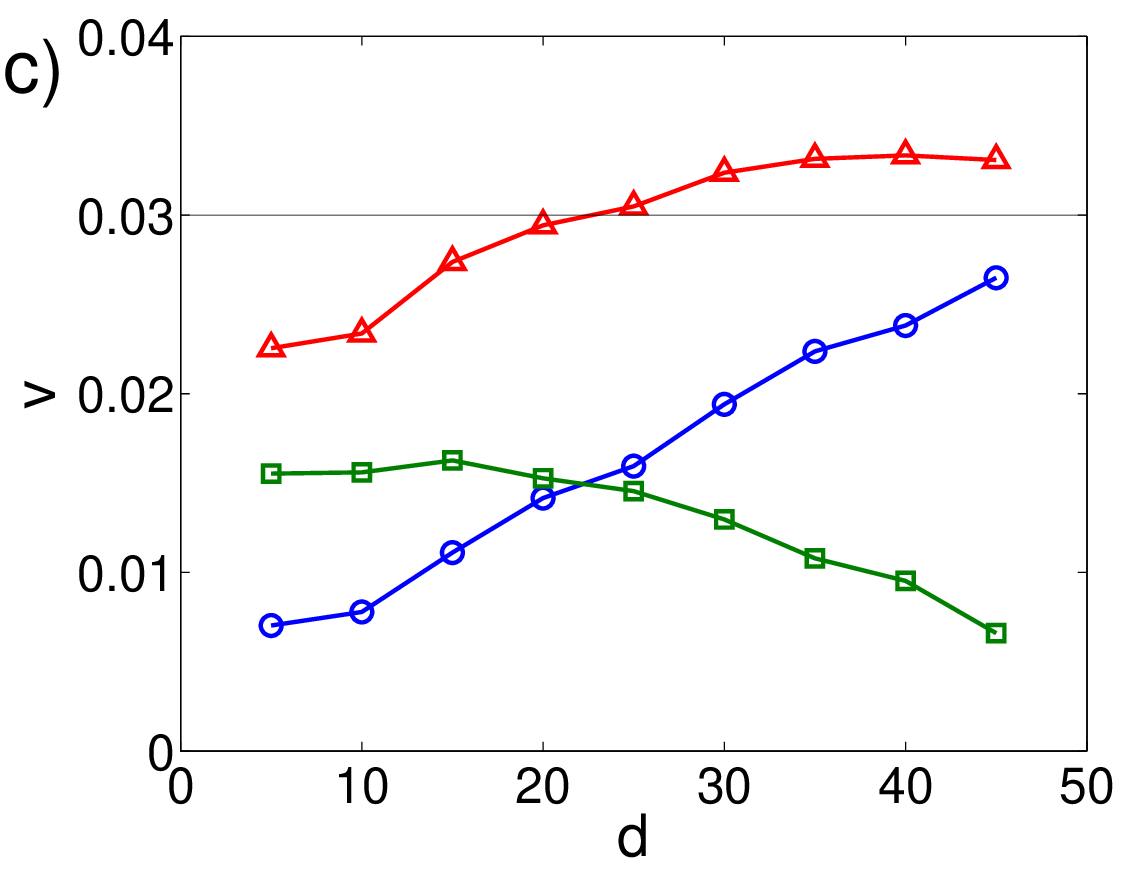} ~ %
\includegraphics[width=5.4cm]{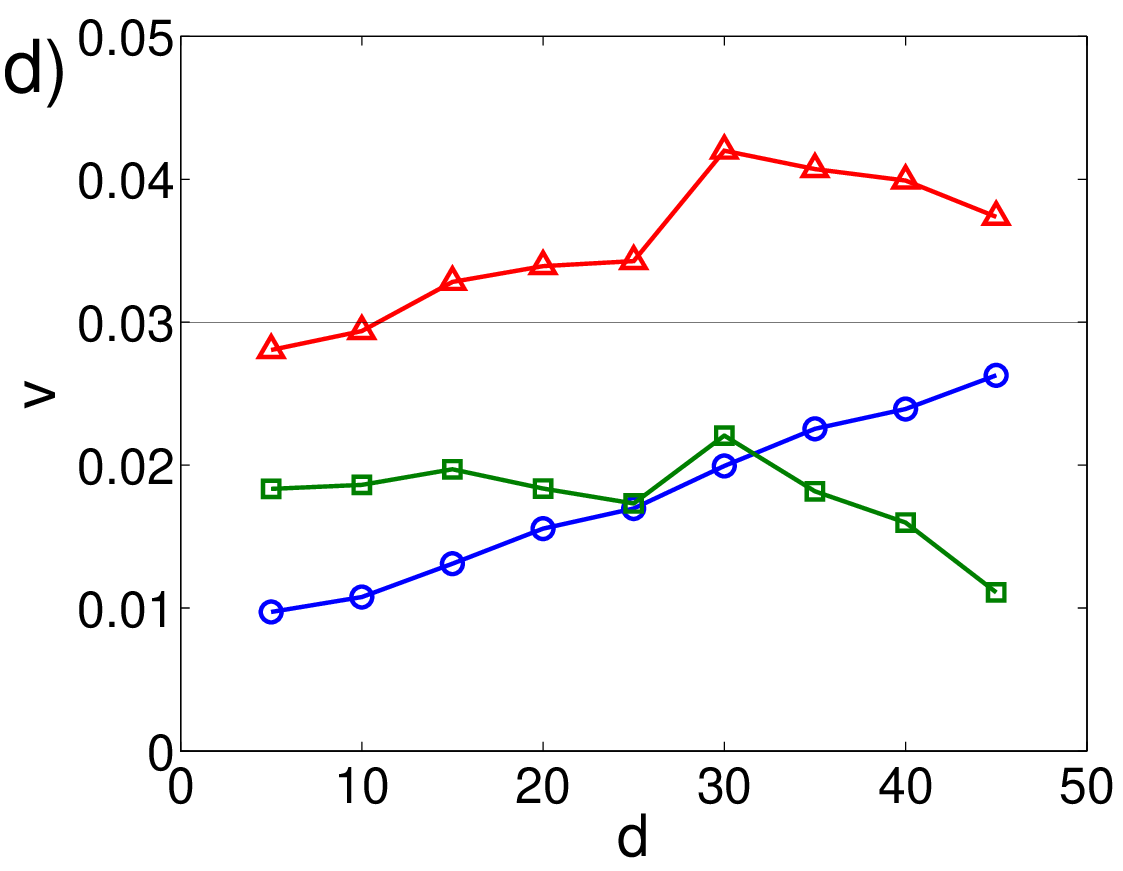}
~ \includegraphics[width=5.4cm]{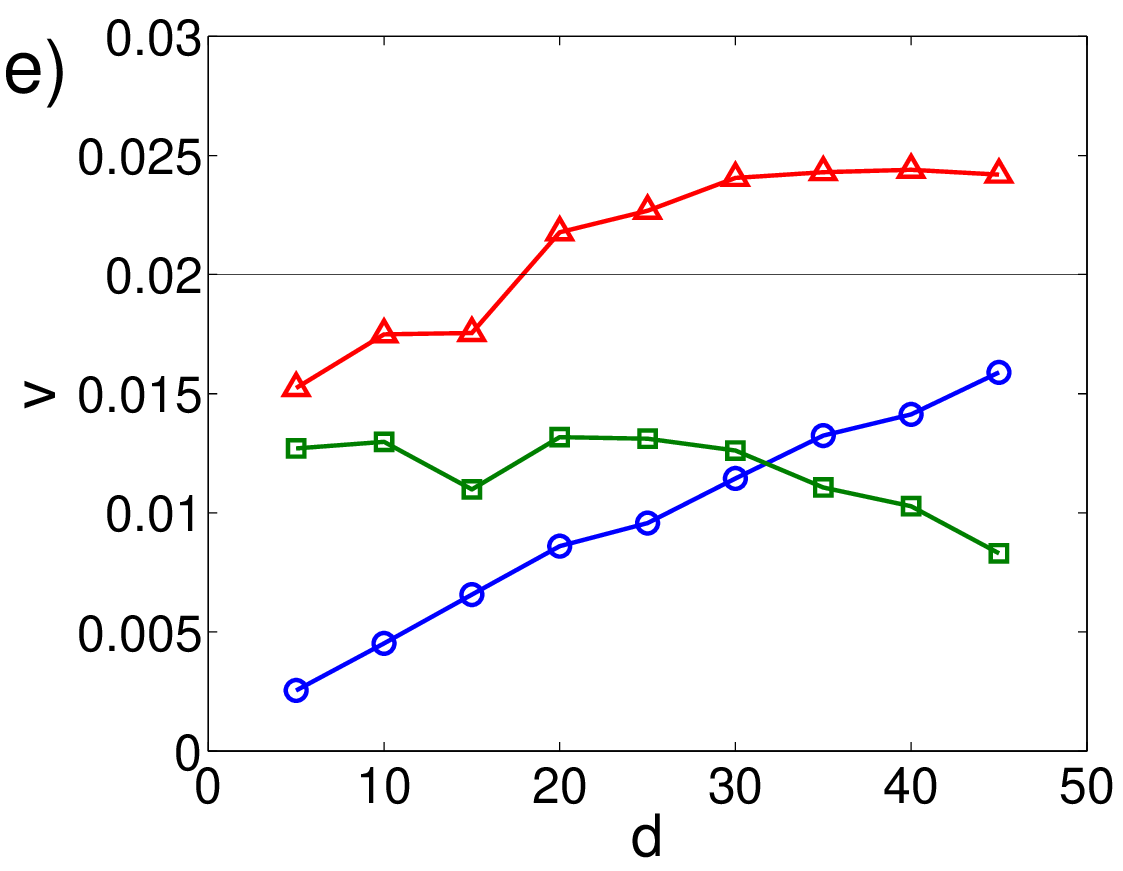}
\end{center}
\vspace{-0.6cm}
\caption{(Color online) Final (post-collision) velocities for the scattering
of two vortices corresponding to the panels depicted in Fig.~\protect\ref%
{scattering}. Depicted in (blue) circles and (green) squares are,
respectively, the final velocities, $v_{1}$ and $v_{2}$, of the initially
moving and quiescent vortices. (Red) triangles depict the sum of these two
velocities, $v_{1}+v_{2} $. The thin horizontal black line designates the
initial velocity of the moving vortex.}
\label{scat_vels}
\end{figure*}

\begin{figure*}[tbh]
\begin{center}
\includegraphics[width=14cm]{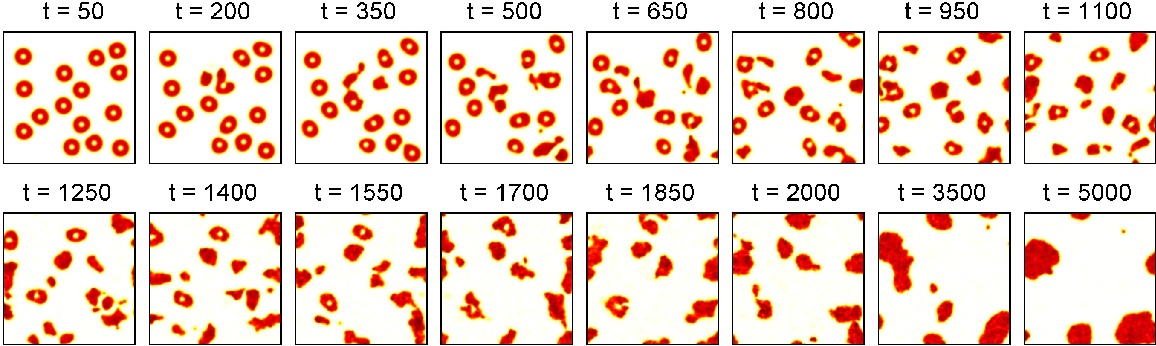}
\end{center}
\par
\vspace{0.0 cm}
\par
\begin{center}
\includegraphics[width=14cm]{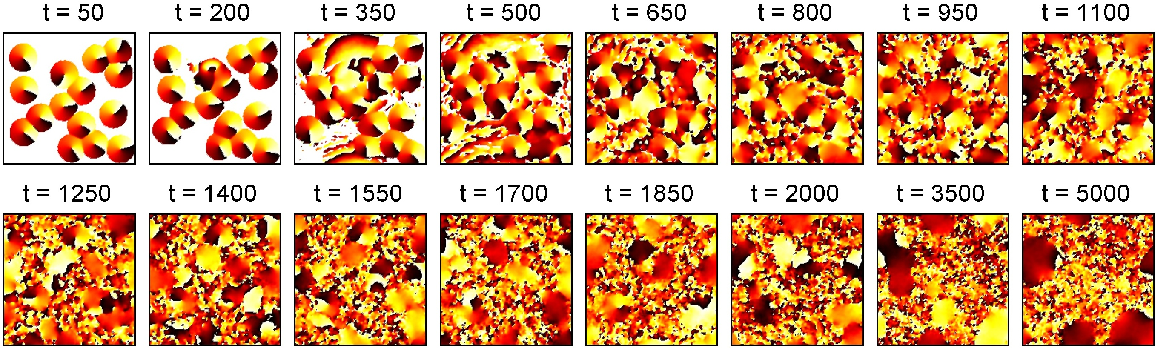}
\end{center}
\vspace{-0.3cm}
\caption{Interaction of a random gas of vortices of charge $m=1$ in a
periodic box. The initial separation between any two neighboring vortices
exceeds twice the width of a single vortex. The initial phases are set
randomly. The first two rows depict the evolution of the field intensity at
the times indicated, while the bottom two rows depict the respective phase
of the field. Note the coarsening effect due to vortex merger and
destruction.}
\label{Nvecs}
\end{figure*}

Clearly, even when the repulsive force acts between vortices, but the
collision velocity is sufficiently large, the vortices have enough momentum
for causing them to merge and eventually break up as well. This case is
depicted in the third column of Fig.~\ref{collision_series}.

Slowly moving vortices with opposite phases at the point of contact are
quite robust against the collisions, as illustrated in the fourth column of
Fig.~\ref{collision_series}. In this panel we depict a ``billiard"-type
example, in which one vortex collides with another, which in turn collides
with a third one. All collisions in this case are elastic because the
relative phases have been chosen such that the colliding sides are always
out-of-phase, providing for the necessary repulsion. It is worthy to note
that, although these collisions at low velocities seem to be elastic, the
stable vortices clearly have internal breathing modes, that are excited
during the collisions. Therefore, a small fraction of the collisional energy
is transferred to these internal modes, preventing the collisions from being
completely elastic. A more in-depth analysis of the excitation of these
internal modes and of the degree of the inelasticity of the collisions fall
outside of the scope of the present work. Nonetheless, in order to quantify
the degree of the elasticity of the collisions, we have measured the
critical velocity $v_{\mathrm{crit}}$ for the vortices to elastically bounce
off each other: at $v<v_{\mathrm{crit}}$ the vortices bounce elastically,
while at $v>v_{\mathrm{crit}}$ they merge and, most often, get destroyed.
The crucial parameter that controls $v_{\mathrm{crit}}$, for a particular
combination of the charges of colliding vortices, is their phase difference $%
{\Delta }${$\phi $. Different values of }$\Delta \phi ${\ allow to have
different relative phases at the colliding sides (as discussed above), and
thus (partially) repel or attract. This effect is illustrated by Fig.~\ref%
{vcrit}, where we depict $v_{\mathrm{crit}}$ versus }${\Delta }${$\phi $ for
vortex pairs with charges (a) $(m_{1},m_{2})=(1,1)$, (b) $%
(m_{1},m_{2})=(1,2) $, and (c) $(m_{1},m_{2})=(2,2)$. From panel (a) it is
clear that, for charges $m_{1}=m_{2}=1$, the most ``stable" (in terms of
observing elastic collisions at larger collisional velocities) collision
occurs around ${\Delta \phi =0}$. This corresponds exactly to the most
stable case mentioned above, when the collisional sides are out of phase. In
panel (b) we depict $v_{\mathrm{crit}} $ for $(m_{1},m_{2})=(1,2)$. It is
seen in this panel that $v_{\mathrm{crit}} $ is $\pi $-periodic, as a
function of }$\Delta \phi $,{\ due to the angular symmetry of the $m_{2}=2$
vortex. For this vortex-charge combination, the most stable collision occurs
around }${\Delta }${$\phi =\pi /2$ (or, due to the symmetry, }${\Delta }${$%
\phi =3\pi /2$), which again corresponds to the adjacent out-of-phase sides.
Finally, in panel (c) we present the results for two vortices with $%
m_{1}=m_{2}=2$. The behavior is very similar to that in panel (b): the same
periodicity, shape and value of the optimal phase difference. However, since
the vortices with $m=2$ vortices are less stable than those with $m=1$, the
critical velocity $v_{\mathrm{crit}}$ is lower for the $(m_{1},m_{2})=(2,2)$
charge combination than for the case of $(m_{1},m_{2})=(1,2)$.}

{In order to fully characterize the collisions between vortices with
different charges, we studied the scattering of a vortex of charge $m_{1}$
impinging at velocity $v_{1}^{\mathrm{ini}}$ upon another vortex of charge $%
m_{2}$ at rest ($v_{2}^{\mathrm{ini}}=0$), with impact parameter $d$. The
impact parameter $d$ is defined as the perpendicular distance between the
trajectory of the incoming vortex and the initially quiescent one. In Fig.~%
\ref{scattering} we show several orbits for the two vortices with different
impact parameters, vortex-charge combinations, and incoming velocities. It
is seen, in all the panels, that all the trajectories are well defined for
the chosen parameters, and clearly feature elastic scattering. The key to
achieve elastic scattering is to choose, for each charge combination, the
optimal phase difference, using Fig.~\ref{vcrit} [i.e.,~$\Delta \phi =0$ for
$m_{1}=m_{2}=1$ and }${\Delta }${$\phi =\pi /2$ for $(m_{1},m_{2})=(1,2)$, $%
(2,1)$, or $(2,2)$], \emph{and} the initial velocity lower than the
corresponding $v_{\mathrm{crit}}$. For example, we chose in panel (a) of
Fig.~\ref{scattering} $\Delta \phi =0$ and $v_{2}^{\mathrm{ini}}=0.03<v_{%
\mathrm{crit}}(\pi =0)=0.067$. In fact, when one uses a velocity closer to
the critical one the vortices start experiencing strong deformations of the
quadrupole and octupole types when they collide nearly head-on. This effect
is clearly visible for the trajectory with $d=5$ (also $d=10$ and $d=20$) in
panel (b) of Fig.~\ref{scattering}, where $v_{2}^{\mathrm{ini}}=0.06$ is
close to the critical value, $v_{\mathrm{crit}}=0.067$, and the trajectory
features undulations due to the internal deformation of the vortex. Even
larger velocities, i.e.,~above $v_{\mathrm{crit}}$, lead to elastic
scattering for larger values of the impact parameter $d$, and to
annihilation of the colliding vortices at smaller $d$ (not shown here). It
is interesting to note that, as for hard spheres with different masses, the
scattering of vortices can also produce very different scattering angles,
depending on the mass combination of the two colliding objects. This effect
is made evident by the comparison of the scattering between vortices with $%
m_{1}=1$ and $m_{2}=2$ in panel (c), and the scattering between $m_{1}=2$
and $m_{2}=1$ in panel (d) (all the other parameters coincide in both
panels). When a ``lighter" vortex with $m_{1}=1$ collides with the
``heavier" one, with $m_{2}=2$, for $d\leq 15$, the former vortex bounces
back after the collisions. In the reverse situation, with the ``heavier" $%
m_{2}=2$ vortex impinging upon the ``lighter" one with $m_{1}=1$, the
``heavier" vortex only loses a portion of its momentum and continues to move
in a straighter trajectory, while the ``lighter" vortex is pushed away at a
larger velocity.}

To further characterize the scattering between vortices we analyzed the
scattering angle as a function of the impact parameter. In our scattering
experiments, we initially define horizontally moving and quiescent vortices.
After the collision, the trajectory of the moving vortex is deviated by
scattering angle $\theta _{1}$ (in our case negative since we consider the
moving vortex placed \emph{below} the quiescent one), and the initially
quiescent vortex is scattered at angle $\theta _{2}$ (in our case,
positive). In Fig.~\ref{scat_angles} we depict the scattering angles
corresponding to the scattering experiments displayed in Fig.~\ref%
{scattering}. We depict by (blue) circles the negative of the scattering
angle, $-\theta _{1}$, of the incoming vortex, and by (green) squares the
scattering angle $\theta _{2}$. In all the cases, we also depict the
difference between the scattering angles, $\Delta\theta =\theta _{2}-\theta
_{1}$, by (red) triangles. In the case of the elastic collision between two
hard spheres of equal masses it is well known that $\Delta\theta=90^{\mathrm{%
o}}$. Figure~\ref{scat_angles} reveals that the collisions between vortices
with equal charges [$m_{1}=m_{2}=1$ in panels (a) and (b), and $%
m_{1}=m_{2}=2 $ in panel (e)] produce $\Delta\theta $ very close to ${90}^{%
\mathrm{o}}$ for all values of the impact parameter. In contrast, as it is
the case for hard spheres with different masses, our simulations demonstrate
that $\Delta\theta > 90^{\mathrm{o}}$ for $m_{1}<m_{2}$ [$%
(m_{1},m_{2})=(1,2) $ in panel (c)], and that $\Delta\theta <90^{\mathrm{o}}$
for $m_{1}>m_{2}$ [$(m_{1},m_{2})=(2,1)$ in panel (d)].

{Another measure of the scattering can be given in terms of the initial and
final velocities. In Fig.~\ref{scat_vels}, we present the final velocities
corresponding to the scattering events shown in Fig.~\ref{scattering}. The
figure shows, by (blue) circles and (green) squares, the post-collision
velocities, }$v_{1}$ and $v_{2}$,{\ respectively, of the initially moving
and quiescent vortices. We also depict, using (red) triangles, the sum of
the final velocities, $v_{1}+v_{2}$, and the thin horizontal black line
represents the initial velocity of the moving vortex, $v_{1}^{\mathrm{ini}}$%
. It can be concluded that $v_{1}+v_{2}<v_{1}^{\mathrm{ini}}$ for relatively
small values of the impact parameter, }$d$, {and $v_{1}+v_{2}>v_{1}^{\mathrm{%
ini}}$ for larger values of }$d$.

The above results suggest that the interactions between vortices may be
either elastic or destructive, depending on the parameters. In order to
simulate the interplay of these interactions in a multi-vortex gas, we
display in Fig.~\ref{Nvecs} a time series depicting the time evolution of a
random vortex gas. The first two rows depict the local intensity plots at
the indicated moments of time, while the bottom two rows depict the
corresponding field phases. We use a square domain of size $120\times 120$
with periodic boundary conditions, seeding 15 vortices of charge $m=1$ at
random positions (such that initially no two vortices were closer than twice
the size of a single vortex), and with random phases. As seen from the
figure, due to the random phase differences and collision angles, some
vortex pairs experience elastic collisions, while others tend to merge and
destroy each other. Since the destruction of vortices is an irreversible
process, all the vortices get eventually destroyed, leading to a seemingly
disordered pattern of interacting intensity patches with an approximately
constant phase inside each one (see the correspondence between field patches
in the top rows and the phase patches in the bottom rows, at the late stage
of the evolution). The dynamics resembles the grain coarsening in typical
chaotic spatiotemporal systems, see e.g., review \cite{sand}.


\section{Conclusions}

In this work, we have revisited the issues of the existence, stability, and
interactions of solitary vortices in the two-dimensional CQNLS
(cubic-quintic nonlinear Schr{\"{o}}dinger) equation, using both numerical
and analytical methods. The latter one is based on the variational approach,
that has been shown to be increasingly more accurate as the vortex'
topological charge increases. We have also developed the analysis of the
azimuthal modulational (in)stability of vortices, using the approximation
originally proposed in other contexts in Refs.~\cite{Koby} and \cite{NLSxMI}%
, which postulates the separation between the frozen radial profile of the
vortex soliton and free evolution of azimuthal perturbations. This approach
leads to an effectively one-dimensional equation for the azimuthal dynamics.
Examining the stability of the perturbed annulus within the framework of the
latter equation, we were able to predict the stability of the vortices
against the breakup. We then ran full 2D simulations to verify the
semi-analytical predictions.

For azimuthally unstable vortices, our predictions of the largest
growth rate are fairly accurate over a wide range of the vortex'
intrinsic frequency (especially for the higher-order vortices, with
topological charge $m>2$). The so developed semi-analytical
technique may be useful for other applications. The analytical
approach is less accurate in predicting the stability border
(intrinsic critical frequency of the vortex) for $m>1$. This
discrepancy is due to deviation from the assumption admitting the
separation of the perturbed evolution into radial and azimuthal
parts, while the VA proper, which was used to predict the radial
shape of the stationary vortex solitons turns out to yield accurate
results. Comparing our numerical results for the critical frequency
with previously published ones, we have
concluded that the most accurate findings were reported in Ref.~\cite%
{AMIxCQNLSxNEW}. We have also found that higher-order vortices may be
subject to a snaking radial instability.

The semi-analytical approach developed in this work may be quite useful for
other 2D models ---at least, for the description of the most stable vortex
solitons with lowest values of the topological charge.

We have also investigated, in detail, collisions between stable vortex
solitons in the two-dimensional CQNLS equation. The outcome of the collision
crucially depends on the phase difference and relative velocity of the
vortices. We have produced the results showing the critical velocity for
elastic collisions as a function of the relative phase for all combinations
of the vortices with $m=1$ and $2$. We also studied the dependence on the
scattering angles and post-collision velocities on the impact parameter of
the collision.

One relevant direction for the extension of the present analysis is to find
accurate conditions for the stabilization of the nearly isotropic vortex
solitons by an external spatially periodic (lattice) potential, see Ref.~\cite%
{Sakaguchi} and references therein. A challenging problem is to extend the
semi-analytical considerations to \emph{three-dimensional} solitons with
embedded vorticity, in the model with the same cubic-quintic nonlinearity.
In fact, it is not known if such 3D vortex solitons with higher vorticity ($%
m\geq 2$) may be stable \cite{PRL,Torner}.

\section{Acknowledgments}

The authors thank Juan Belmonte and Jes\'us Cuevas for providing useful insights. RCG
gratefully acknowledges support from NSF-DMS-0806762. PGK gratefully
acknowledges support from NSF-DMS-0349023 (CAREER) and NSF-DMS-0806762, as
well as from the Alexander von Humboldt Foundation.

\end{document}